\documentclass{aastex631}
\usepackage{gensymb}
\usepackage{hyperref}
\usepackage[caption=false]{subfig}
\usepackage [autostyle, english = american]{csquotes}
\MakeOuterQuote{"}

\begin{document}

\title{Scattering Parameter Measurements of the Long Wavelength Array Antenna and Front End Electronics}

\correspondingauthor{Christopher DiLullo}
\email{cdilullo@cpi.com}

\author[0000-0001-5944-9118]{Christopher DiLullo}
\affiliation{NASA Goddard Space Flight Center \\
Observational Cosmology Laboratory, Code 665 \\
8800 Greenbelt Rd. \\
Greenbelt, MD 20771}

\author[0000-0003-2661-0598]{Whitham D. Reeve}
\affiliation{Reeve Engineers \\
2211 Kissee Ct. \\
Anchorage, AK 99517}

\author[0000-0003-1996-3141]{Brian C. Hicks}
\affiliation{Naval Research Laboratory \\
Rapid Prototyping and Engineering Section, Code 7212 \\
4555 Overlook Ave. SW \\
Washington, DC 20032}

\author[0000-0003-1407-0141]{Jayce Dowell}
\affiliation{University of New Mexico \\ 
210 Yale Blvd. NE \\ Albuquerque, NM 87131}

\begin{abstract}
We present recent 2--port vector network analyzer (VNA) measurements of the complete set of scattering 
parameters for the antenna used within the Long Wavelength Array (LWA) and the associated front end 
electronics (FEEs). Full scattering parameter measurements of the antenna yield not only the reflection coefficient for each 
polarization, S11 and S22, but also the coupling between polarizations, S12 and S21. 
These had been previously modeled using simulations, but direct measurements had not been 
obtained until now. The measurements are used to derive a frequency dependent impedance mismatch factor (IMF) which represents the 
fraction of power that is passed through the antenna--FEE interface and not reflected due to a mismatch between the 
impedance of the antenna and the impedance of the FEE. We also present results from a two antenna experiment where 
each antenna is hooked up to a separate port on the VNA. This allows for cross--antenna coupling to be measured for 
all four possible polarization combinations. Finally, we apply the newly measured IMF and FEE forward gain
corrections to LWA data to investigate how well they remove instrumental effects.
\end{abstract}

\section{Introduction} \label{sec:intro}
In a linear electrical system, the scattering parameters describe how incident power is either 
reflected or transmitted at a boundary. In a  2--port measurement, four scattering parameters can be 
measured by sending a signal from Port 1 to Port 2 and from Port 2 to Port 1: S11, S21, S12, and S22 which describe the reflected power in Port 1, 
the transmitted power from Port 1 to Port 2, the transmitted power from Port 2 back to Port 1, and the reflected power at 
Port 2, respectively. Measuring the S11 and S22 parameters is equivalent to measuring the mismatch in electrical impedance at the 
boundary.

A mismatch between the impedance of any antenna and the impedance of its associated electronics 
will cause some reflection of power. In a receiving system such as a radio telescope, this reflection 
of power corresponds to a loss of sensitivity that is frequency dependent. This frequency dependent loss of 
sensitivity is an issue for experiments which require absolute calibration of the telescope since this effect must 
be quantified in order to set the absolute zero calibration level. However, if the reflection coefficients 
can be directly measured using a Vector Network Analyzer (VNA), then the relative efficiency of the system 
as a function of frequency, known as the impedance mismatch factor \citep[IMF;][]{alan1986}, can be computed. 
The IMF can be applied as a correction to data to remove impedance mismatch effects and can be expressed as
\begin{equation} \label{eq: imf}
    IMF = \frac{(1 - |\Gamma_{ant}|^2) (1-|\Gamma_{RX}|^2)}{|1 - \Gamma_{ant} \Gamma_{RX}|^2} ,
\end{equation}
where $\Gamma_{ant}$ and $\Gamma_{RX}$ are the reflection coefficients of the antenna and receiving electronics, 
respectively.

The Long Wavelength Array (LWA) is a low frequency radio telescope consisting of two stations in New Mexico, USA and
a thrid station in California, USA. 
LWA1 \citep{taylor2012} is colocated with the Karl. J. Jansky Very Large Array (VLA) in Soccorro, New Mexico, 
LWA--SV \citep{cranmer2017} is located on the Sevilleta National Wildlife Refuge in New Mexico, and LWA--OVRO
is located at the Owens Valley Radio Observatory in California. The New Mexico stations 
are antenna arrays consisting of 256 elements with pseudo--random placement while LWA--OVRO is currently being 
upgraded to support 352 elements with a different array geometry. The LWA antenna \citep{hicks2012} consists 
of two perpendicular "blade" dipoles whose geometry was chosen to balance cost, stability, and RF performance. 
The LWA antenna has been deployed by numerous groups all over the world due to its low cost, robust stability,
wideband RF performance, and ease of construction.

\citet{hicks2012} simulated the impedance of the LWA antenna and converted this into an impedance matching 
efficiency (IME) given by 
\begin{equation} \label{eq: ime}
    IME = 1 - |\Gamma_{ant}|^2,
\end{equation}
which is simply Equation \ref{eq: imf} under the assumption that $\Gamma_{RX} = 0$. However, it is unlikely that 
the FEEs have no reflection at all, i.e. the FEE is not perfectly matched to a reference impedance of 50 $\Omega$
that is constant in frequency. 
Thus, actual measurements of both the antenna and FEEs were required for a 
more accurate picture of the impedance mismatch properties of the system. The current LWA experiments requiring 
a better understanding of the impedance mismatch characteristics, are the LWA1 Low Frequency Sky Survey 
\citep[LWA1 LFSS;][]{dowell2017} and the effort to detect the redshifted 21 cm signal from neutral hydrogen present
during the formation of the first stars \citep{dilullo2020, dilullo2021}. The LWA1 LFSS is the major driver 
of the work presented here because an impedance mismatch correction is applied to the raw data at the first step 
of calibration. The resulting absolute temperature calibration of the survey is reliant on this correction and so 
any errors in the correction will follow through the entire calibration pipeline. It is therefore important to 
make actual measurements of the impedance mismatch factor so that the limitations of the model presented in 
\citet{hicks2012} could be better understood and an improved correction could be applied to previous and 
future survey data.

This paper details work carried out in November of 2022 which made direct measurements of the LWA 
antenna scattering parameters at three locations in New Mexico and the scattering parameters of the LWA FEEs, 
which were measured in February of 2023. 
Custom Calibration Fixtures were designed in order to de--embed the FEE from the antenna measurements. 
Cross--antenna coupling measurements were also performed by connecting two nearby antennas to each port of the VNA.
Mutual coupling between antennas will change their electromagnetic properties, such as impedance and gain 
pattern, in a way that is not easy to model due to the shear complexity of the problem and required computation 
time. This is a concern for both the LWA1 LFSS and the 21 cm cosmology experiment at LWA--SV, referenced above, since these 
projects use models of the antenna gain pattern to set their calibration. Attempts have been made to model the 
effects of mutual coupling for a LWA station using method of moments simulations \citep{ellingson2011}; however, actual measurements 
have been impossible until now. The paper is structured as follows:
Section \ref{sec: fixtures} details the custom calibration and test fixtures which were designed for this experiment,
Section \ref{sec: measurements} details the single antenna measurements, the antenna--antenna coupling measurements, and presents
the results from each, and Section \ref{sec: discussion} discusses the results and applies them as new corrections to LWA1 data
to investigate how they capture instrumental effects compared to older corrections.

\section{Custom Calibration and Test Fixtures} \label{sec: fixtures}
Calibration of a VNA requires three calibration standards: Open, Short, and Load which correspond to infinite resistance, 
zero resistance, and $50 \ \Omega$ resistance, respectively. These references are typically not 
difficult to fabricate for use at low frequencies; however, in the case of the LWA antenna it was slightly more difficult due to the $180 \degree$ hybrid 
coupler which exists on the FEE boards. The hybrid coupler acts as a balun between the balanced dipole feedpoints and unbalanced coaxial cables 
that carry the signal to the backend electronics.
See \citet{hicks2012} and associated references for more detailed
information on the LWA FEE board design. 

\begin{figure}
    \centering
    \subfloat[PCB Layout]{%
    \includegraphics[width=0.5\linewidth]{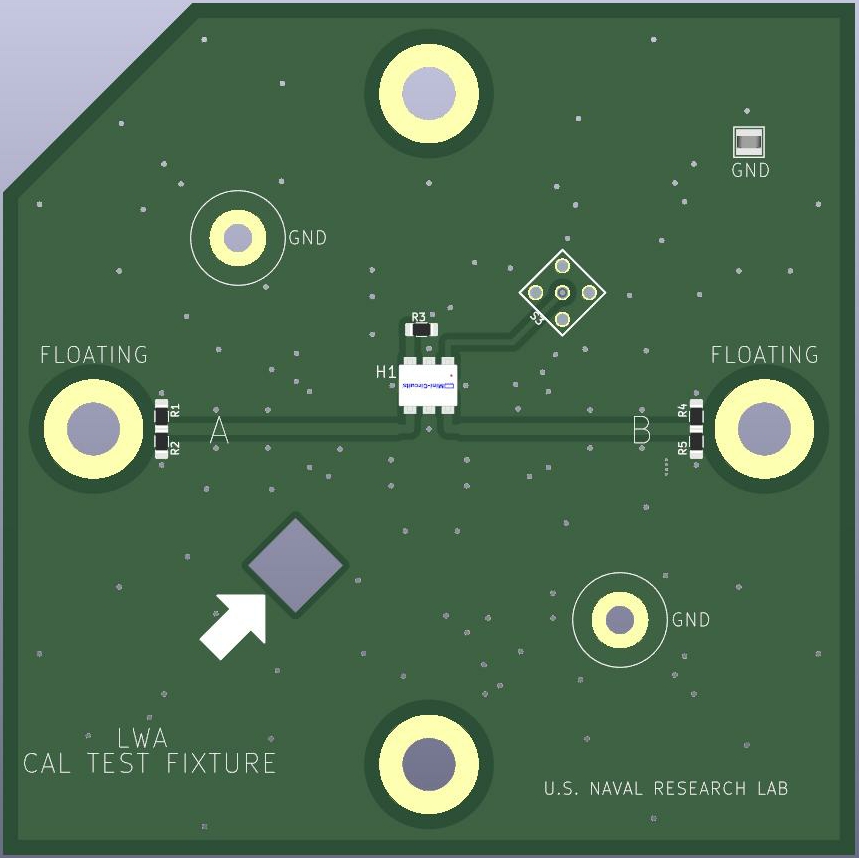}%
    }
    \subfloat[Schematic Diagram]{%
    \raisebox{2.5cm}{\includegraphics[width=0.5\linewidth]{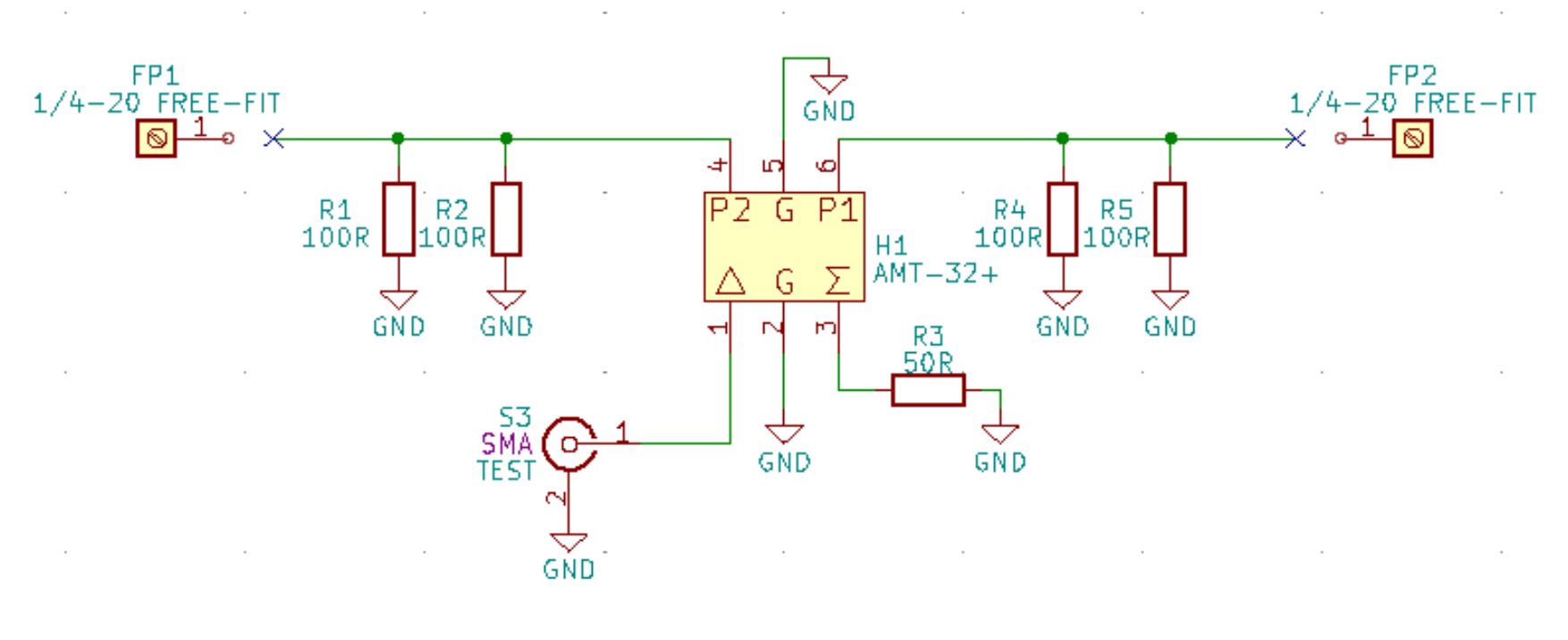}}%
    }
    \caption{Calibration Fixtures. (a) CAD model showing the layout of a Calibration Fixture PCB with the $180\degree$ hybrid coupler shown at the 
    center. The pads at the end of the traces by the circular feed point holes are for the different calibration resistors. The Open standard has 
    no resistors, the Short standard has two $0 \ \Omega$ resistors in parallel, and the Load standard has two $100 \ \Omega$ resistors in 
    parallel. The traces to the connector and hybrid coupler sum load also are visible. The PCB are made from FR4 glass epoxy material and their 
    dimensions are 4.5 in x 4.5 in (114 mm x 114 mm). Each complete Calibration Fixture consists of two PCBs mounted back--to--back. 
    (b) Schematic diagram showing the hybrid coupler at the center, connector, hybrid coupler sum load and the resistor pads at the end of the traces 
    that run to the dipole feedpoints.}
    \label{fig:calibration}
\end{figure}

Custom VNA Calibration Fixtures were designed to include a TeleTech model HX62A hybrid coupler so that its effects could be de--embedded, thus 
moving the measurement reference plane from the VNA to the antenna feedpoints on the antenna hub. The single differential input from the antenna 
feedpoints into the HX62A hybrid coupler is treated as two single ended inputs which are referenced to ground. The impedance of each of these single 
inputs as well as the delta output port is 50 $\Omega$. The Calibration Fixtures use printed circuit 
boards (PCBs) with the same shape and dimensions as the FEE PCBs, but having traces  only to pads near the
dipole feedpoints where the calibration resistors are installed. The pads for the calibration resistors are symmetrical about the traces; thus, a 
Short is achieved by two parallel 0 $\Omega$ resistors, a 50 $\Omega$ Load is achieved by two parallel 100 $\Omega$ resistors, and an Open is 
achieved by not populating the pads with any resistors. The traces from the hybrid coupler do not connect to the feedpoints. The Calibration 
Fixtures also include a coaxial connector and 50 $\Omega$ load resistor for the hybrid coupler sum port. A CAD model and the associated schematic 
diagram of the Calibration Fixtures can be seen in Figure \ref{fig:calibration}. Commercial VNA calibration kits include a set of parameters that 
describe the delays and frequency--dependent coefficients for capacitance and inductance of the calibration standards. These parameters were set to zero 
for the custom calibration standards because the Open, Short, and Load were placed within a millimeter of the feedpoint reference plane, and capacitance 
and inductance effects at the low frequencies of interest are negligible.

Three Calibration Fixtures were fabricated. The three Calibration Fixtures each consist of a pancake of two identical PCBs, in the same manner as the FEE,
which allows the Fixture to connect to both antenna polarizations. A Thru calibration fixture also was fabricated for complete 2--port calibration since 
both ports on the VNA are required for the single antenna and antenna mutual coupling measurements. The Thru Calibration Fixture consists of two Test 
Fixtures, described below, which are connected to each other at the balanced feedpoints and is classified as an Unknown Thru.

\begin{figure}
    \centering
    \subfloat[PCB Layout]{%
    \includegraphics[width=0.5\linewidth]{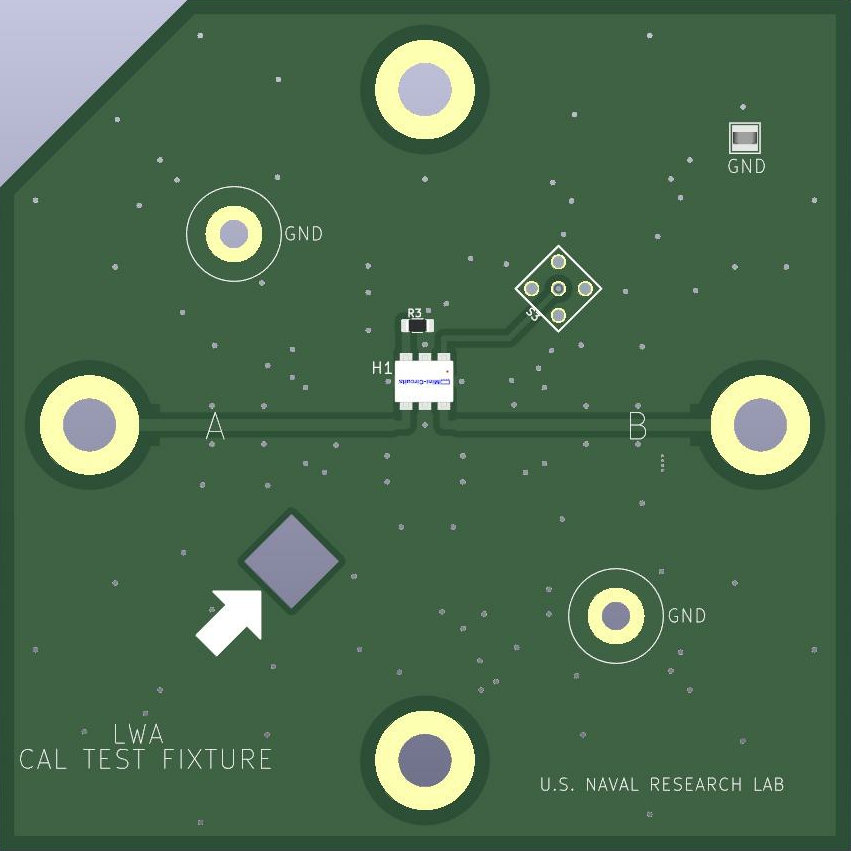}%
    }
    \subfloat[Schematic Diagram]{%
    \raisebox{2.5cm}{\includegraphics[width=0.5\linewidth]{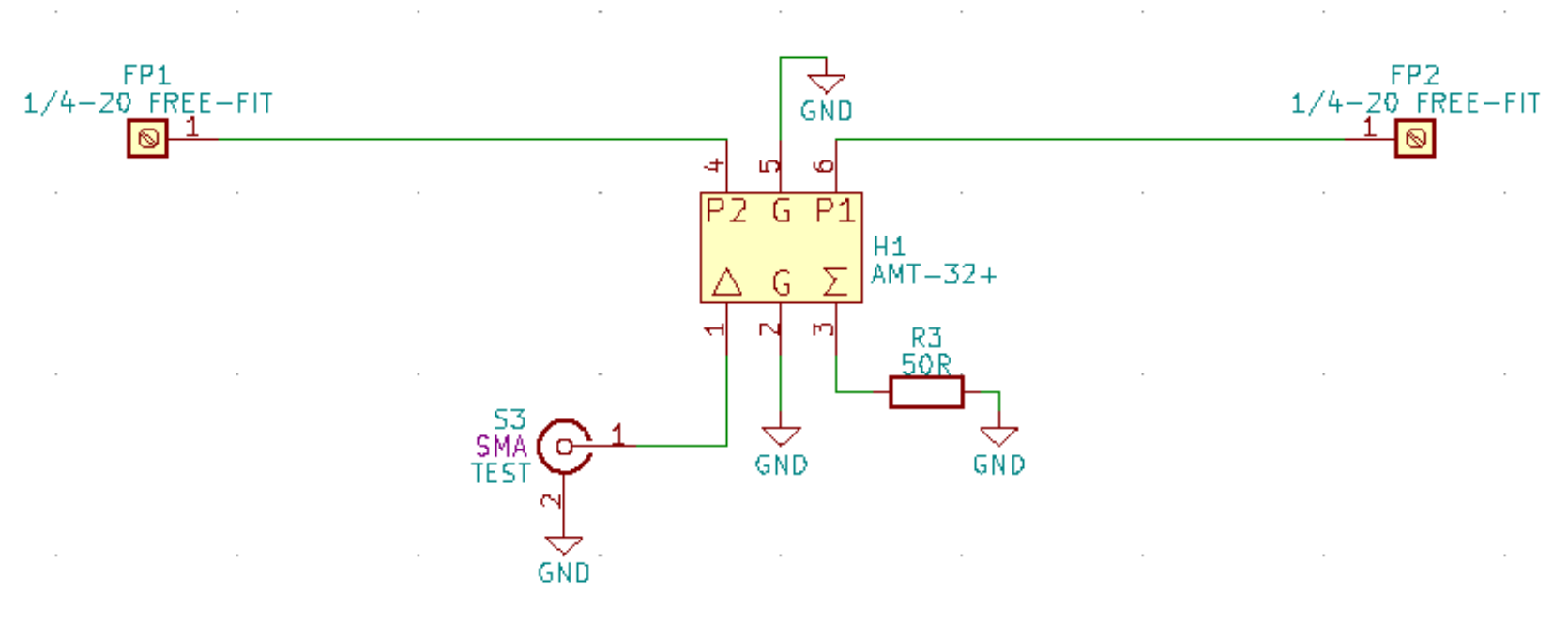}}%
    }
    \caption{Test Fixture. (a) CAD model showing the layout of a Test Fixture PCB. The layout is identical to the Calibration Fixtures except there 
    are no resistor pads and the traces from the hybrid coupler run directly to the feedpoints. (b) Schematic diagram.}
    \label{fig:test}
\end{figure}

In addition to the Calibration Fixtures, a set of Test Fixtures was fabricated to connect the VNA to the antenna--under--test for scattering 
parameter measurements after calibration. The test fixtures had the same dimensions and layout as the Calibration Fixtures and included the hybrid 
coupler; however, they did not include pads for the Open, Short, or Load resistors. Instead, the traces from the hybrid coupler on the 
Test Fixture connects to the dipole feedpoints as seen in Figure \ref{fig:test}. Antenna mutual coupling measurements required a Test 
Fixture on each antenna, so two nominally identical sets of Test Fixtures were fabricated. See Section \ref{sec: antenna} for more details.

\section{Measurements and Results} \label{sec: measurements}
\textit{In situ} scattering parameter measurements were taken between November 10--11$^{\rm{th}}$, 2022 at both commissioned stations in New
Mexico, LWA1 and LWA--SV, as well as an isolated antenna which was setup at a third site at the end of the northern arm of the VLA which will 
soon house another LWA station (LWA--NA). The measurements were divided into two types: single antenna, which measures the scattering parameters
using a single antenna--under--test, and antenna mutual coupling, which uses two test fixtures and two adjacent antennas--under--test in order to 
measure the strength of mutual coupling between adjacent antennas in the array. The following sections discuss the setup and procedures for each 
of the  measurement types. A Keysight Technologies N9917A FieldFox vector network analyzer was used for all antenna measurements. The measurements 
were carried out between 5 and 200 MHz with an IF bandwidth of 10 kHz and spectral resolution of 195 kHz. All measurement procedure documentation, data, and scripts
used in the following analysis are freely available on \href{https://github.com/lwa-project/Antenna_Impedance}{GitHub}\footnote{https://github.com/lwa-project/Antenna\_Impedance}.

\subsection{Single Antenna Measurements} \label{sec: dipole}
\begin{figure}
    \centering
    \subfloat[Single Antenna]{%
    \includegraphics[width=0.5\textwidth]{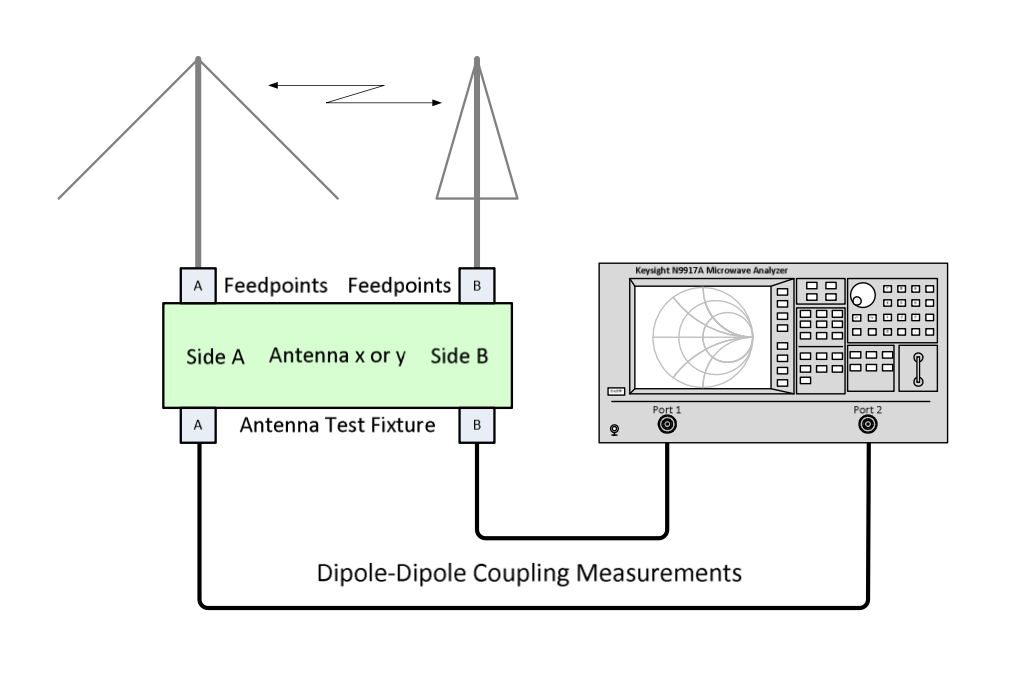}%
    }
    \subfloat[Antenna Mutual Coupling]{%
    \includegraphics[width=0.5\textwidth]{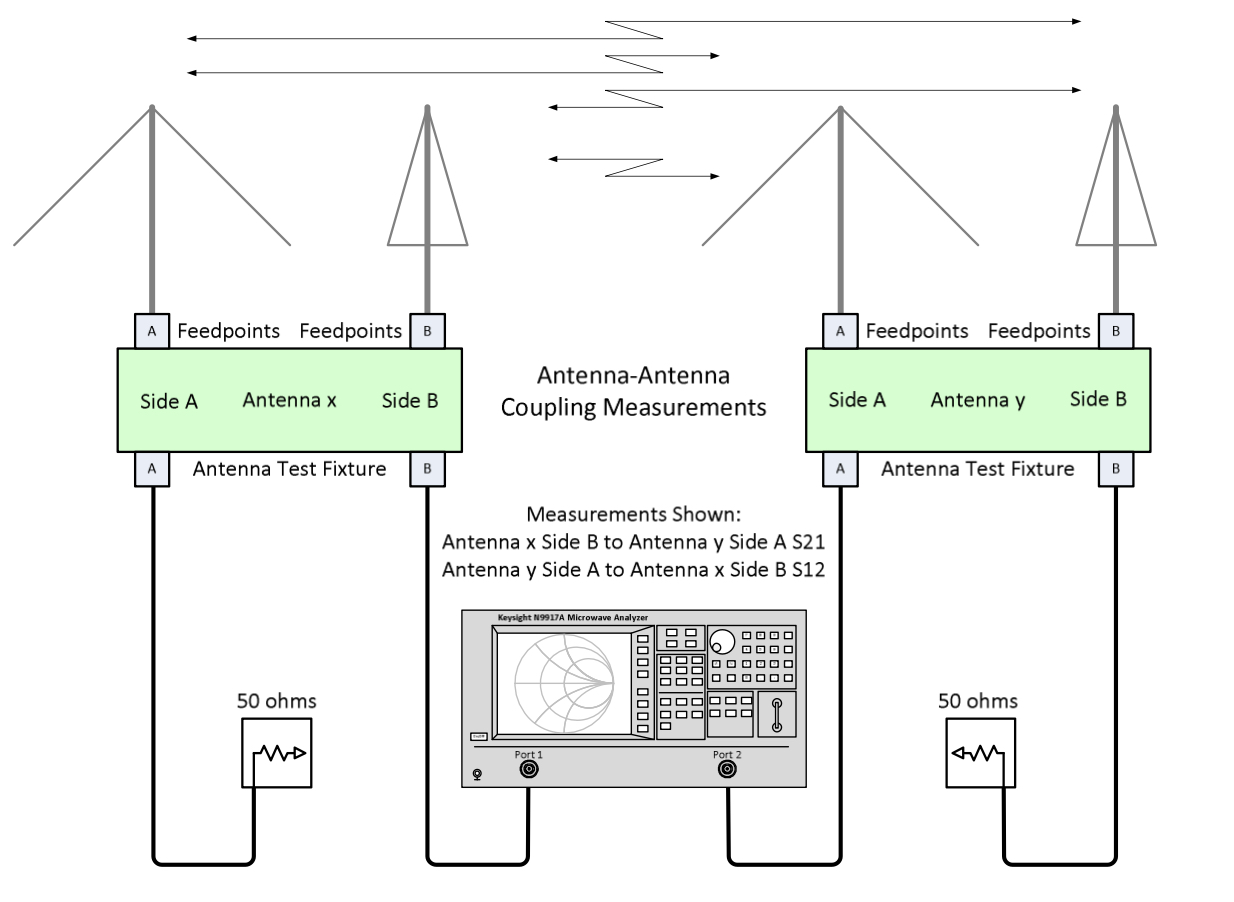}%
    }
    \caption{Example measurement setups for both types of measurements. The thin arrowed lines denote coupling paths that 
    each type of measurement is sensitive to. (a) Single antenna measurements where polarization B of 
    the antenna is connected to Port 1 and polarization A is connected to Port 2. (b) Antenna mutual coupling 
    measurements where polarization B of antenna X is connected to Port 1 and polarization A of antenna Y is 
    connected to Port 2. The inactive dipoles were terminated in 50 ohms via the test cables as shown. 
    Measuring each polarization of each antenna was accomplished by simply reconnecting the cables at the 
    Test Fixture to the desire polarization.}
    \label{fig:setup}
\end{figure}

The single antenna measurements were setup using a single antenna--under--test where the two perpendicular polarizations are connected 
to the two ports on the VNA. See Figure \ref{fig:setup} for an illustrative diagram of the setup. First, the VNA was calibrated using 
the Calibration Fixtures described in Section \ref{sec: fixtures}. A two port calibration of the VNA was carried out where each port 
was calibrated independently using the same calibration standards and two 11 m cables. The cables were laid as close as possible to 
existing antenna cables in order to best replicate how the antenna is usually setup. An Unknown Thru was used in the last calibration step to connect Port 1 to 
Port 2 once both ports were independently calibrated.

The Test Fixture described in Section \ref{sec: fixtures} was then used to make the scattering parameter measurements. 
These measurements not only allow for reflection parameters to be measured for the independent dipole polarizations via measurements 
of S11 and S22, it also allows for measurements of the cross--coupling between the individual dipole arms via measurements of S12 and S21. 
We denote this coupling as "dipole--dipole coupling", which should not be confused with mutual coupling between separate antennas, 
discussed in Section \ref{sec: antenna}. Measurements were made at different antenna locations within the array to capture any 
location--dependent effects that might change the response of antennas deeply embedded within the array.

\begin{figure}
    \centering
    \subfloat[LWA1]{%
    \includegraphics[width=\textwidth]{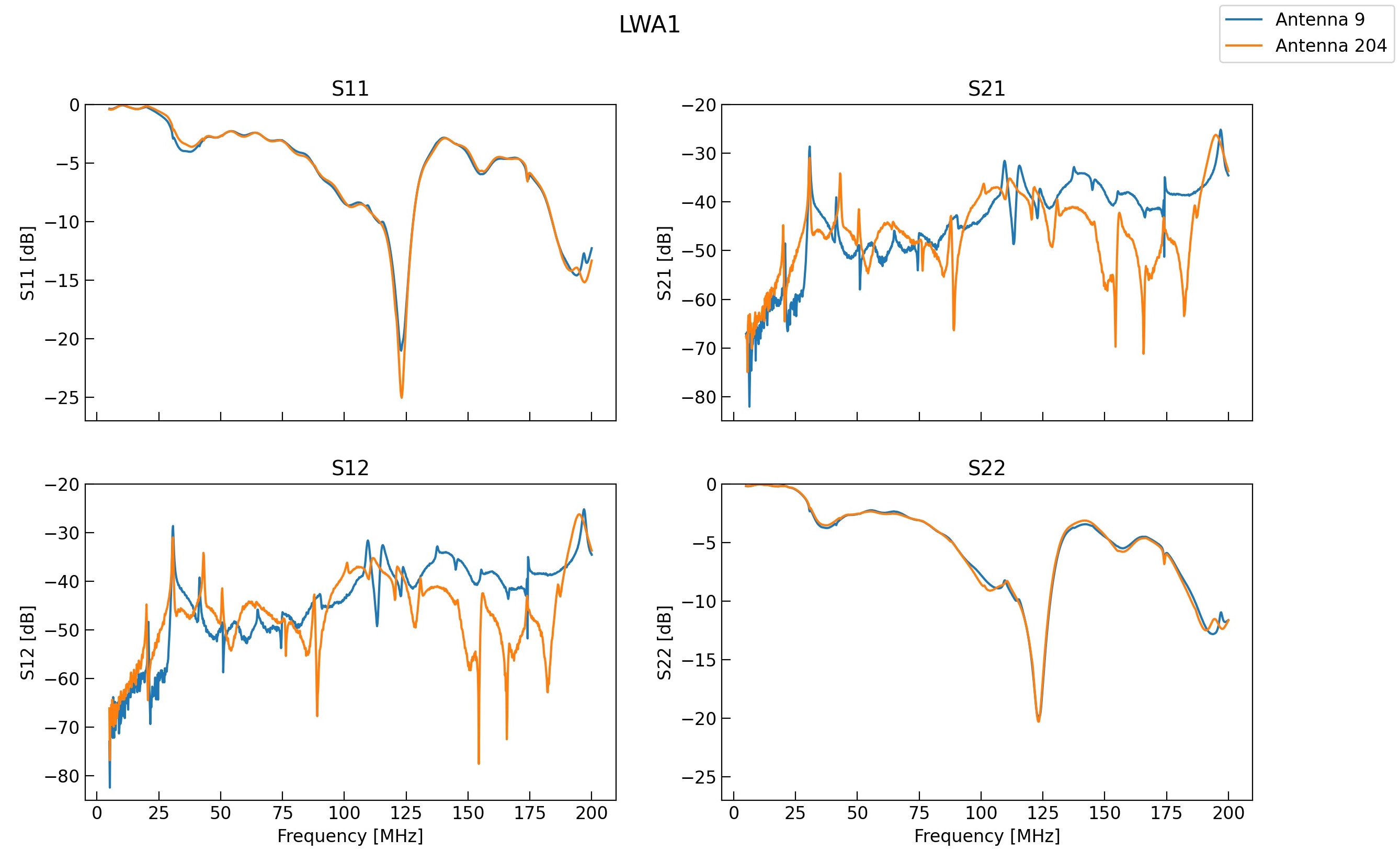}%
    } \\
    \subfloat[LWA--SV]{%
    \includegraphics[width=\textwidth]{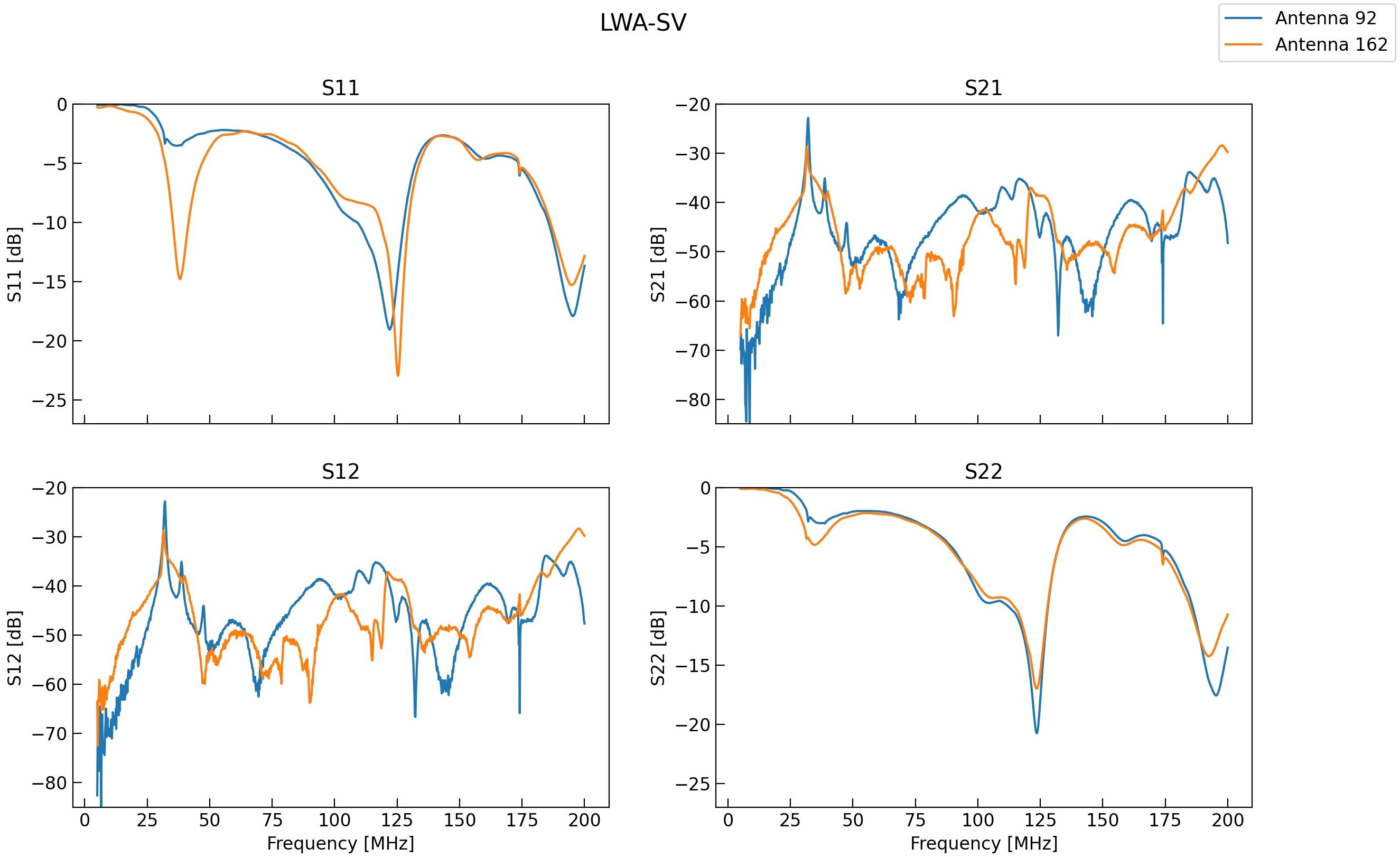}%
    }
    \caption{Single antenna scattering parameter measurements at the two completed LWA stations in New Mexico.
    (a) LWA1 measurements where antenna 9 is deeply embedded within the array and antenna 204 is on the edge.
    (b) LWA--SV measurements where antenna 92 is deeply embedded within the array and antenna 162 is on the edge.}
    \label{fig:s_params}
\end{figure}

The results for LWA1 and LWA--SV are shown in Figure \ref{fig:s_params}. We find that the two pairs: S11 and S22; and S21 and 
S12, for each measured antenna, are generally reciprocal. This is encouraging to see despite the lack of symmetry
in the antenna environments. Differences between the quality of each polarization arm or their environments, e.g. variances 
in construction or varying distances to nearby antennas, can result in variances between S11 and S22, or S21 and S12; however, 
we only find variance on the order of the noise level of the measurements. It is also apparent from Figure \ref{fig:s_params}
that the variances between scattering parameters across antennas is small, which again would not be generally expected due to
variances in their respective environments within the array. This implies that it is a reasonable assumption that different 
antennas across the array have similar impedance characteristics.

We note a deep resonance around 125 MHz, but in the typical LWA observing band between 10--88 MHz the typical amplitudes of 
S11 and S22 are between -3 dB and -5 dB. However, S11 and S22 approach 0 dB, i.e. the antenna becomes almost entirely reflective, at 
frequencies $\nu \lesssim 25$ MHz which implies the sensitivity of the array degrades at these frequencies. The presence of a ripple 
of unknown origin can be seen in the S11 and S22 measurements of order $\sim 10$ MHz at both LWA1 and LWA--SV. Ripple--like structures can arise from 
many types of calibration errors that might be present during field measurements such as these. A few examples of such an errors would be drifts in 
cable performance due to temperature variations throughout the measurements and possible cable reflections due to a mismatch between the source 
impedance and cable impedance. However, we believe the former scenario to be unlikely as the VNA was calibrated immediately before the 
measurements were made at each antenna and the ambient temperature was relatively stable over the course of a single measurement. Cable reflections 
could possibly explain the ripple since the cables where 11 m long LMR--240 cables with a velocity ratio, $v / c = 0.84$. This could cause a ripple
with a characteristic scale of 11.4 MHz, which is close to the $\sim 10$ MHz ripple seen; however, we again believe our calibration procedure would 
have captured an effect such as this.

We see a deep resonance in S11 for antenna 162 at LWA--SV centered around 38.5 MHz 
with an amplitude of -15 dB, but do not see such an extreme feature in the measurements of other antennas. We conclude this 
feature is anomalous since the other measurements agree that the amplitude of this dip is more on the order of -3.5 dB. 
The amplitudes of S12 and S21, which measure the amplitude of cross--coupling between antenna polarizations, are very small with 
averages of -46 dB. This means that we do not expect any appreciable amount of polarization leakage which can 
degrade the sensitivity of the antenna and the station as a whole. 

\begin{figure}
    \centering
    \includegraphics[width=\linewidth]{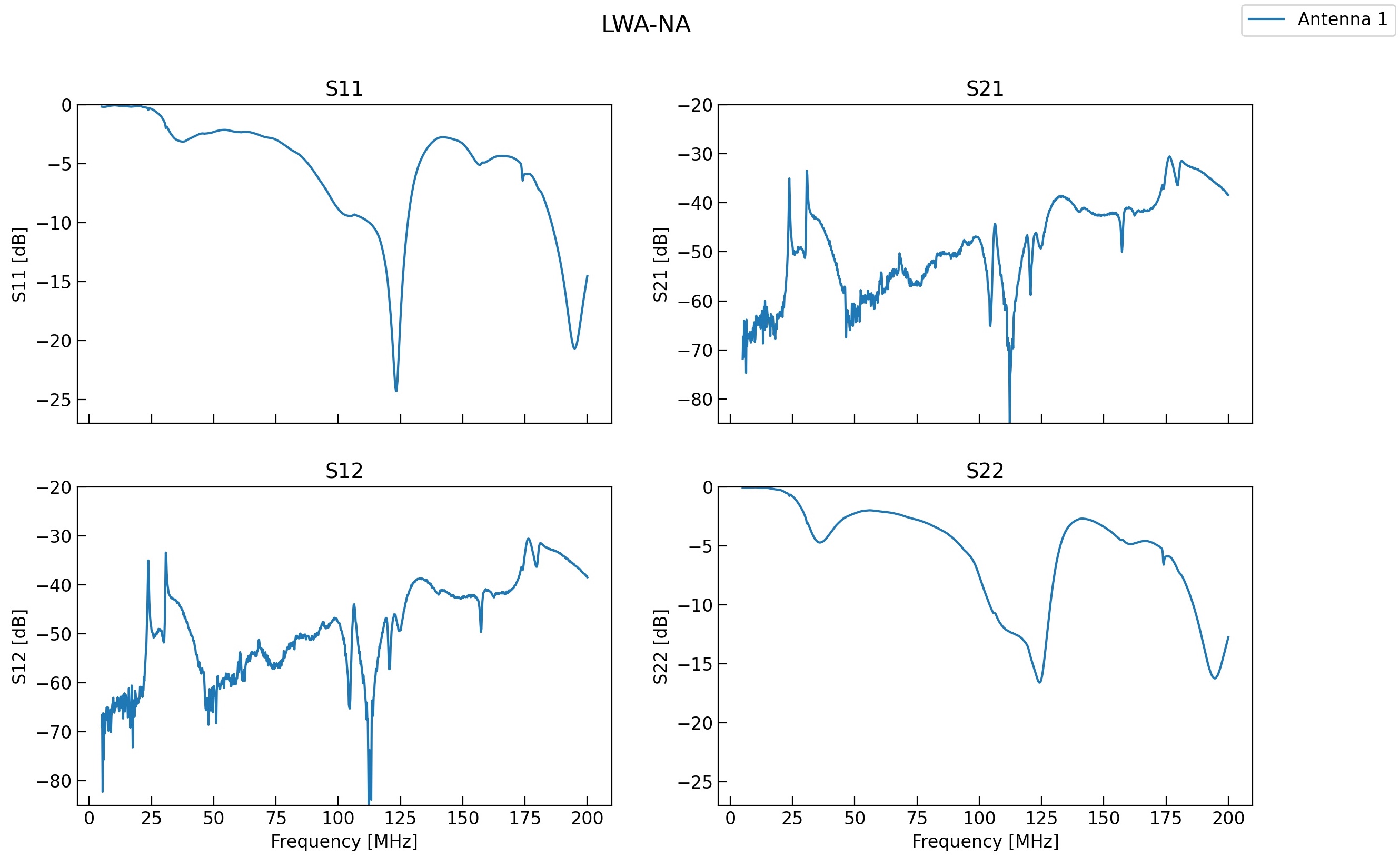}
    \caption{Scattering parameter measurements for a single antenna built at the future LWA--NA site.
    This is the most isolated measurement of a LWA antenna with regard to the presence of other 
    antennas.}
    \label{fig:s_params_na}
\end{figure}

The results for an isolated antenna are shown in Figure \ref{fig:s_params_na}. This was a single antenna setup at 
the future LWA--NA site and therefore represents the most isolated measurement of a LWA antenna. The measurements at both 
LWA1 and LWA--SV suffer from some degree of antenna--antenna mutual coupling, but this was avoided at LWA--NA since no other 
antennas there have been constructed. We find good agreement between the measurements taken at LWA--NA and the other stations.

\subsection{Antenna Mutual Coupling} \label{sec: antenna}
The antenna mutual coupling measurements used two adjacent antennas in the array to measure how strongly 
one polarization of the first antenna couples to a polarization of the second. The setup can be seen in 
Figure \ref{fig:setup}. We denote the two antennas in the setup as Antenna X and Antenna Y and the two 
polarizations on each as A and B. Therefore, there are a total of 4 permutations that were measured 
in order to capture the full coupling. Whichever polarization that was not being tested on each antenna was 
connected to a 50 $\Omega$ termination to isolate that polarization so it will not affect
the coupling results.

\begin{figure}
    \centering
    \subfloat[LWA1 Antenna Positions]{%
    \includegraphics[width=0.9\textwidth]{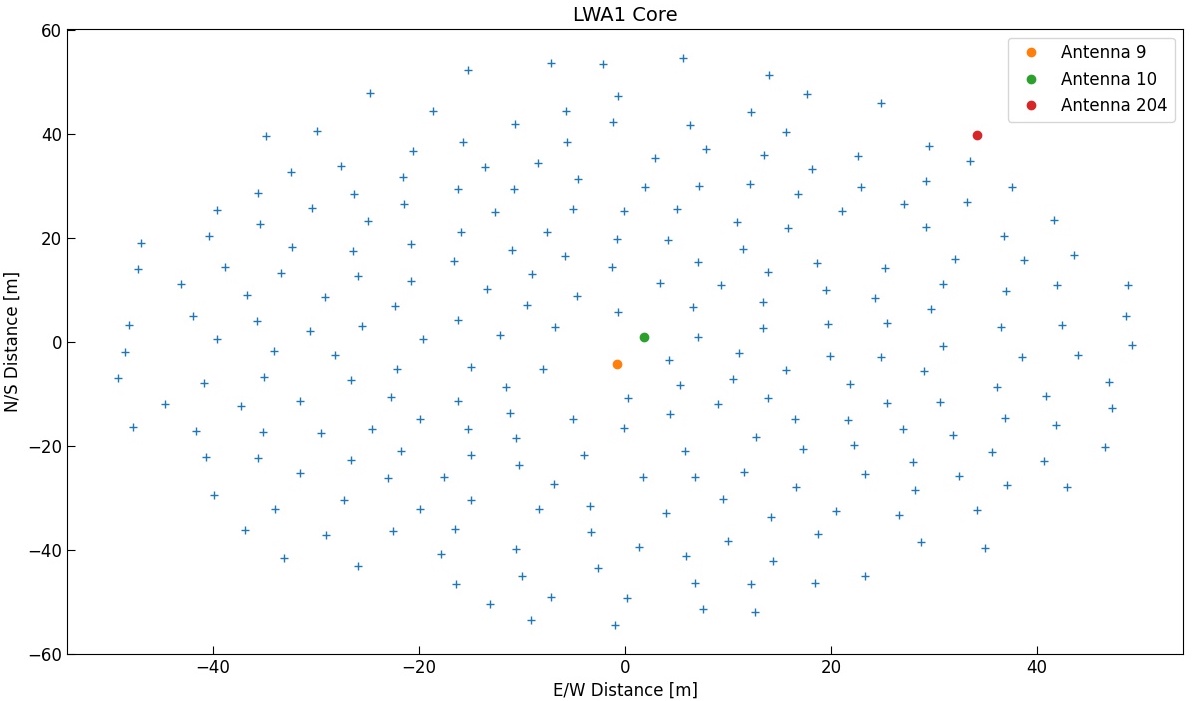}%
    } \\
    \subfloat[S12 Measurements]{%
    \includegraphics[width=0.9\textwidth]{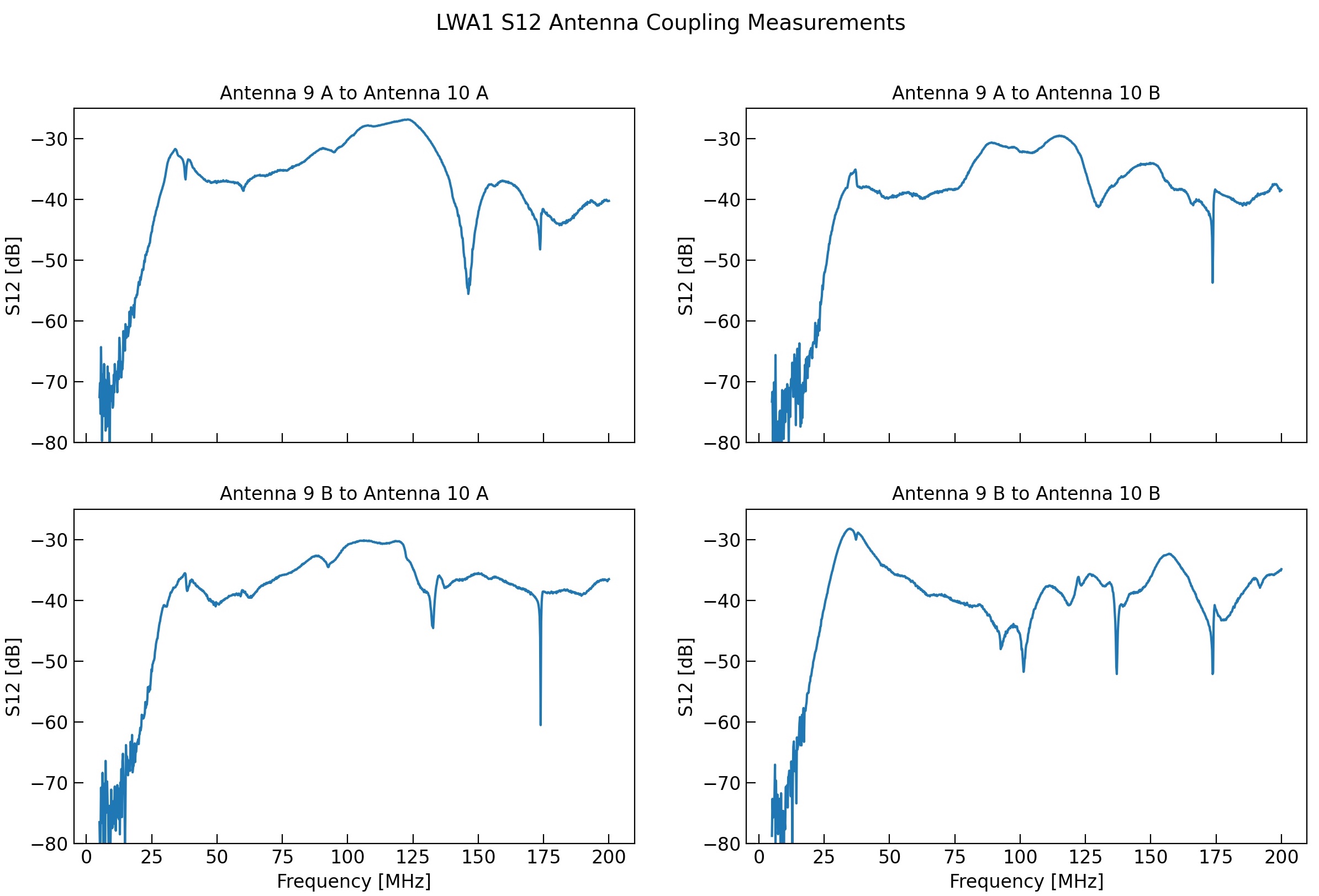}
    }
    \caption{LWA1 Antenna Mutual Coupling Results. (a) Antenna positions within LWA1 with the measured
    antennas marked. Antenna 204 is on the North--Eastern edge of the array and was used for the single antenna measurements. 
    Antennas 9 and 10 are embedded in the array and used for antenna mutual coupling measurements. 
    (b) S12 results for the four possible polarization combinations.}
    \label{fig:lwa1_mutual}
\end{figure}

\begin{figure}
    \centering
    \subfloat[LWA--SV Antenna Positions]{%
    \includegraphics[width=0.9\textwidth]{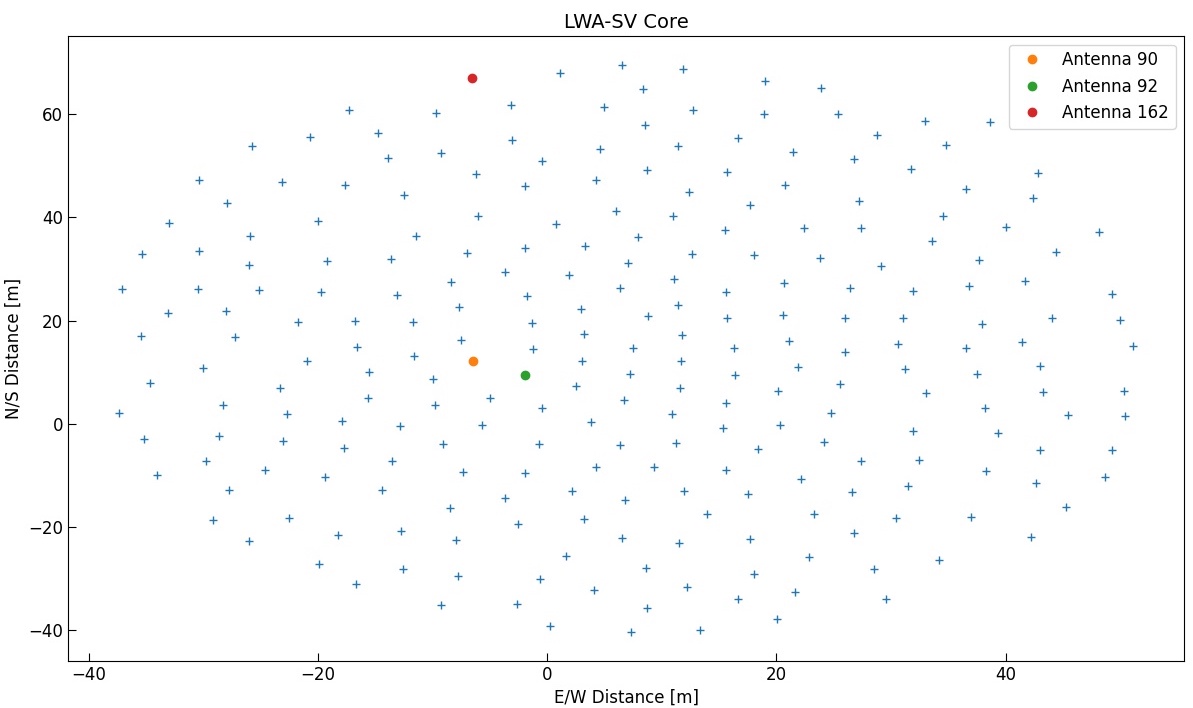}%
    } \\
    \subfloat[S12 Measurements]{%
    \includegraphics[width=0.9\textwidth]{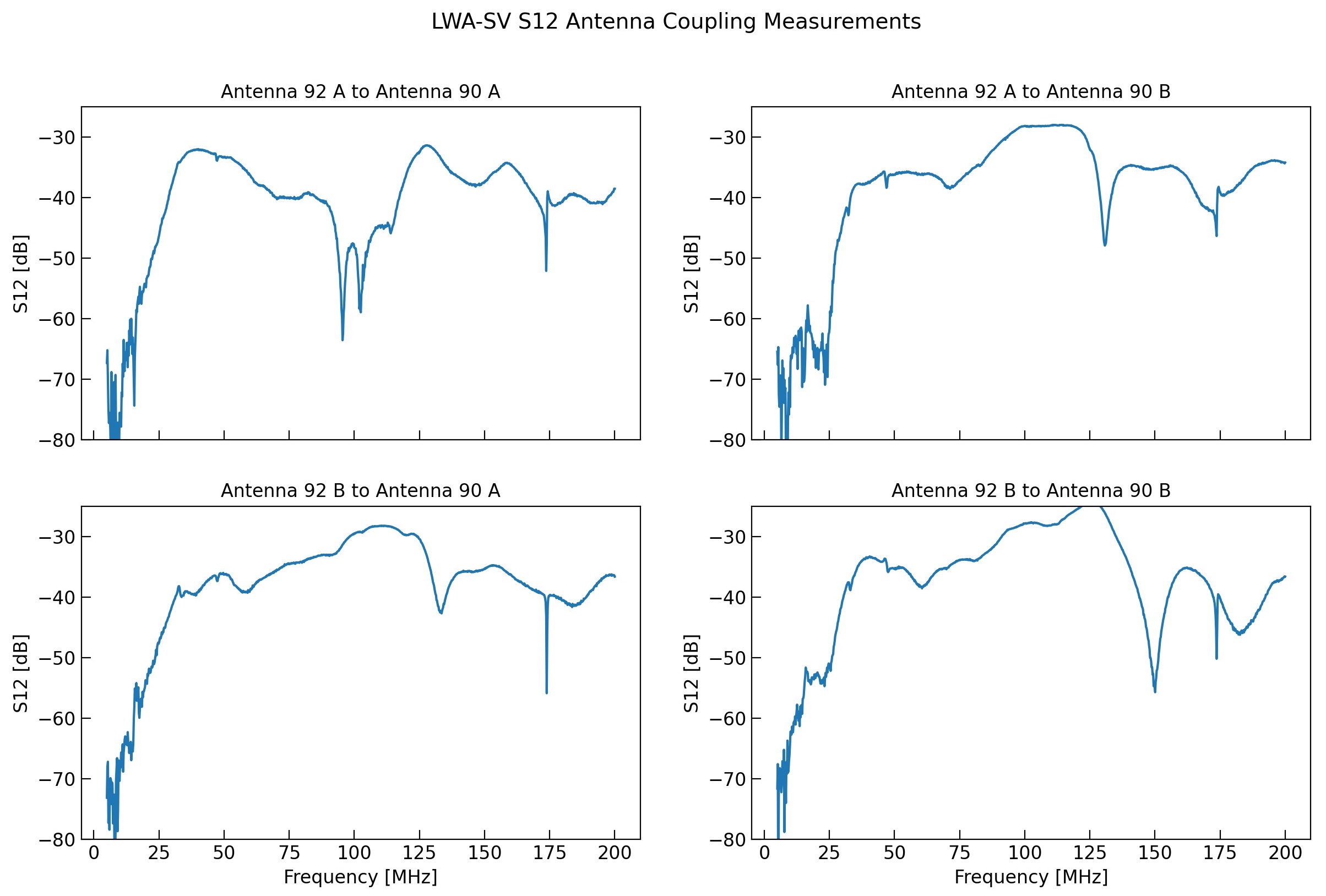}%
    }
    \caption{LWA--SV Antenna Mutual Coupling Results. (a) Antenna positions within LWA--SV with the measured
    antennas marked. Antenna 162 is on the Northern edge of the array and was used for the single antenna measurements. 
    Antennas 90 and 92 are embedded in the array and used for antenna mutual coupling measurements. 
    (b) S12 results for the four possible polarization combinations.}
    \label{fig:lwasv_mutual}
\end{figure}

The two ports of the VNA were independently calibrated in the same manner as the single antenna measurements.
The Thru calibration was achieved by using the Thru Calibration Fixture, see Section \ref{sec: fixtures},
to connect Port 1 to Port 2. The cables which were not terminated to 50 $\Omega$ were pulled through the individual 
antenna masts and connected to the Thru Calibration Fixture in the middle of the two antennas.
Two Test Fixtures were used to essentially turn one antenna into a transmitter and the other into a receiver.
The minimum spacing between two antennas in a LWA array is 5.0 m in order to reduce mutual coupling effects in
real observations. We chose antennas 9 and 10 at LWA1, which are separated by a distance of 5.9 m, and 
antennas 90 and 92 at LWA--SV, which are separated by a distance of 5.2 m. The scattering parameters S12 and 
S21 measure the coupling strength between the two antennas for a given polarization combination and their measurements
are totally reciprocal. Figures \ref{fig:lwa1_mutual} and \ref{fig:lwasv_mutual} show the locations of the 
measured antennas within the arrays and the measured values of S12 for each of the polarization combinations 
at LWA1 and LWA--SV, respectively.

These results are limited in their ability to capture the full effect of mutual coupling between antennas since it
is dependent on all the antennas in the array and is highly nonlinear. The amplitude of mutual coupling will vary 
between antennas and is likely dependent on the location within the array; however, these measurements should 
yield a sense of the general amplitude of the coupling. We find that the mutual coupling between antennas is small with the 
largest observed coupling in the LWA observable frequency range of 3--88 MHz being on the order of -30 dB. These data 
could be useful in developing a statistical model describing how mutual coupling of this amplitude can perturb the synthesized station 
beam pattern, but that is left for future work.

\subsection{Impedance Mismatch Factor} \label{sec: imf}
The major goal of the work presented here was to use the scattering parameter measurements to calculate a new impedance 
mismatch factor, see Equation \ref{eq: imf}, that would better capture the impedance mismatch between the antenna and the 
FEEs than previous electromagnetic simulations of both could. Therefore, in order to properly compute the IMF, the scattering 
parameters for the FEE boards had to be measured also. There are currently two versions of the LWA FEE board: the version 1.8,
which has been in use for the past few years, and the version 2.0, which is newly developed with many improvements over the version 1.8.
A paper detailing the version 2.0 FEE is in development and the boards are just now becoming available, so the following work uses
the version 1.8 FEE board.

\begin{figure}
    \centering
    \subfloat[FEE Test Fixture]{%
    \includegraphics[width=0.33\textwidth]{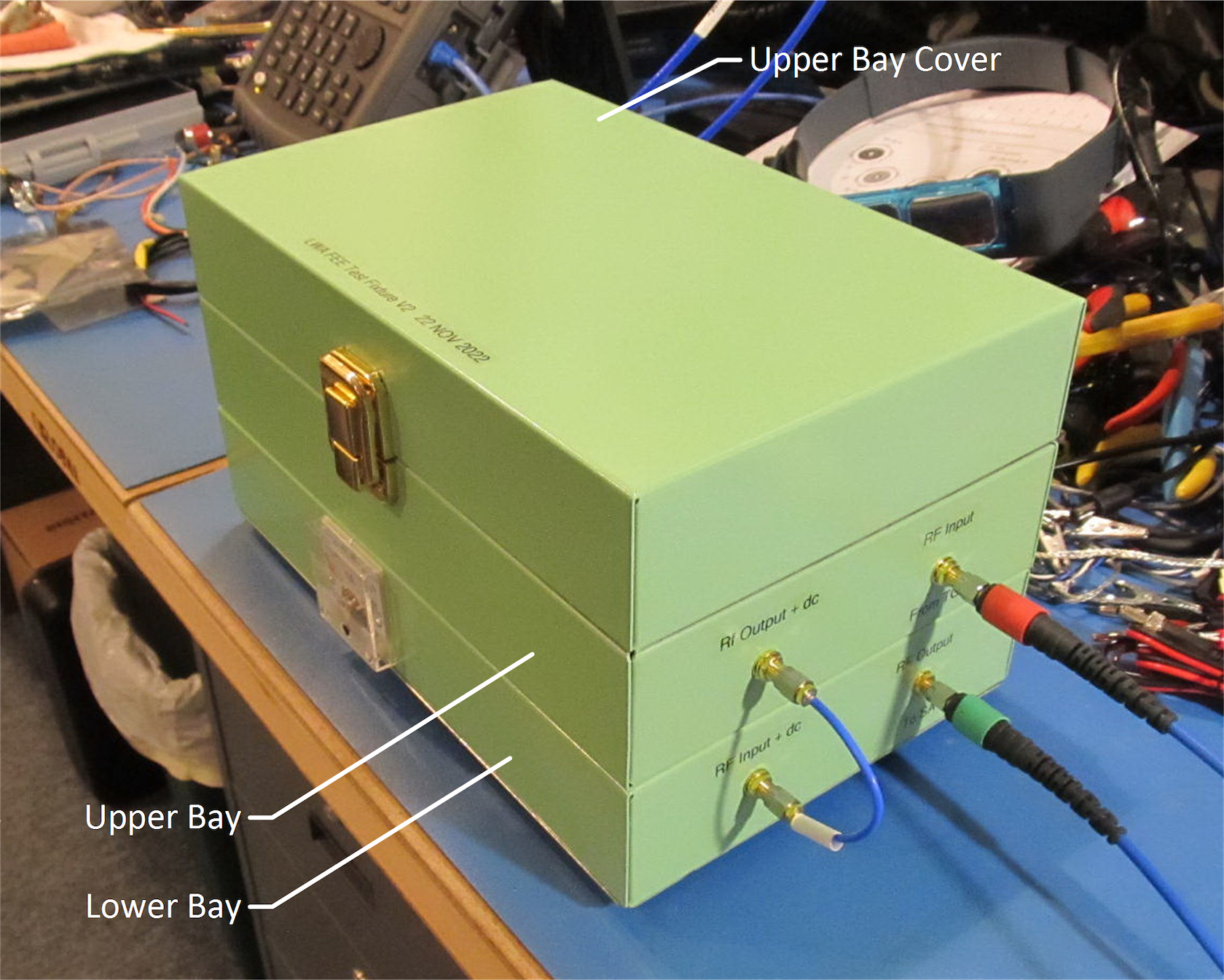}
    }%
    \subfloat[Upper Bay]{%
    \includegraphics[width=0.33\textwidth]{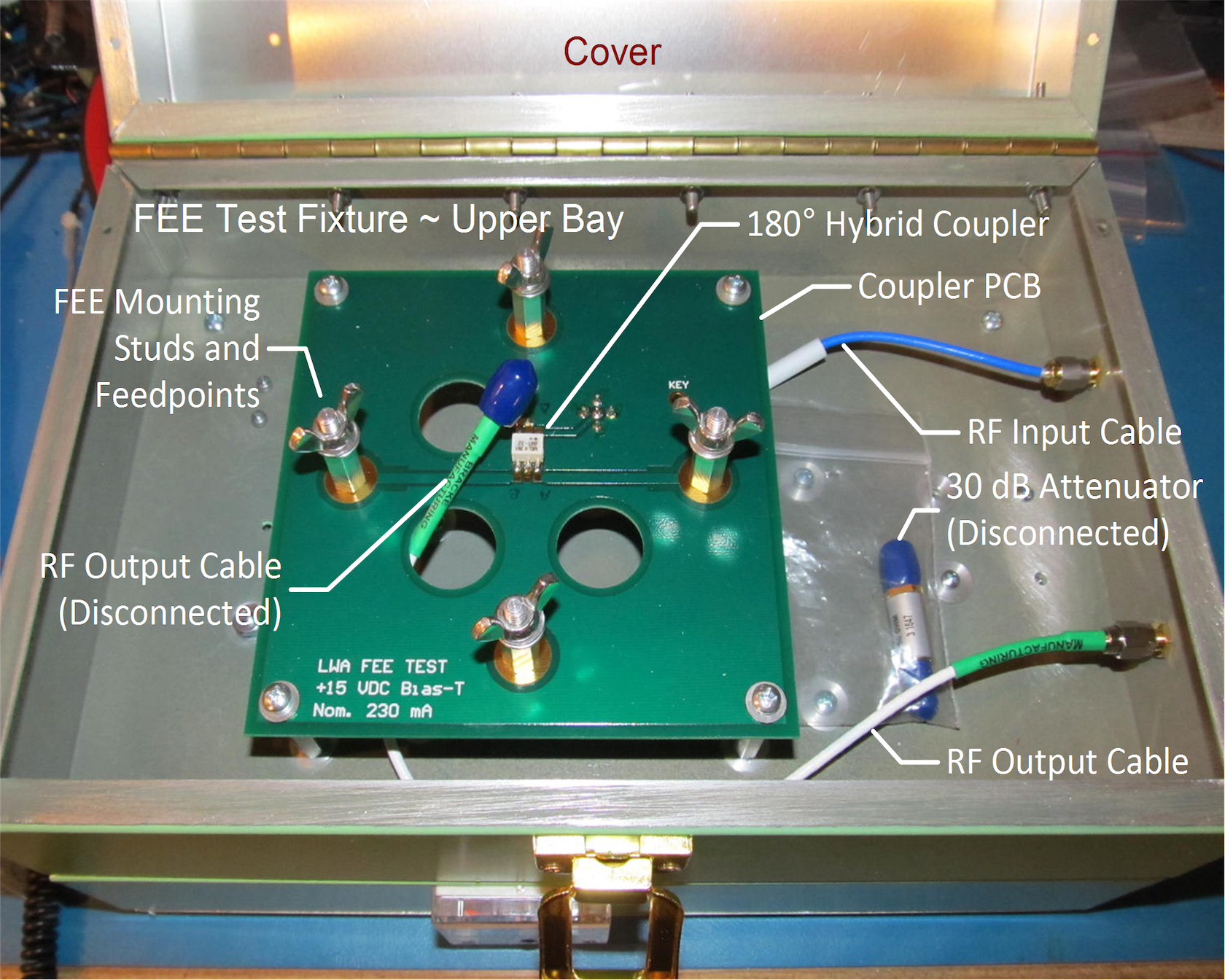}
    }%
    \subfloat[Lower Bay]{%
    \includegraphics[width=0.33\textwidth]{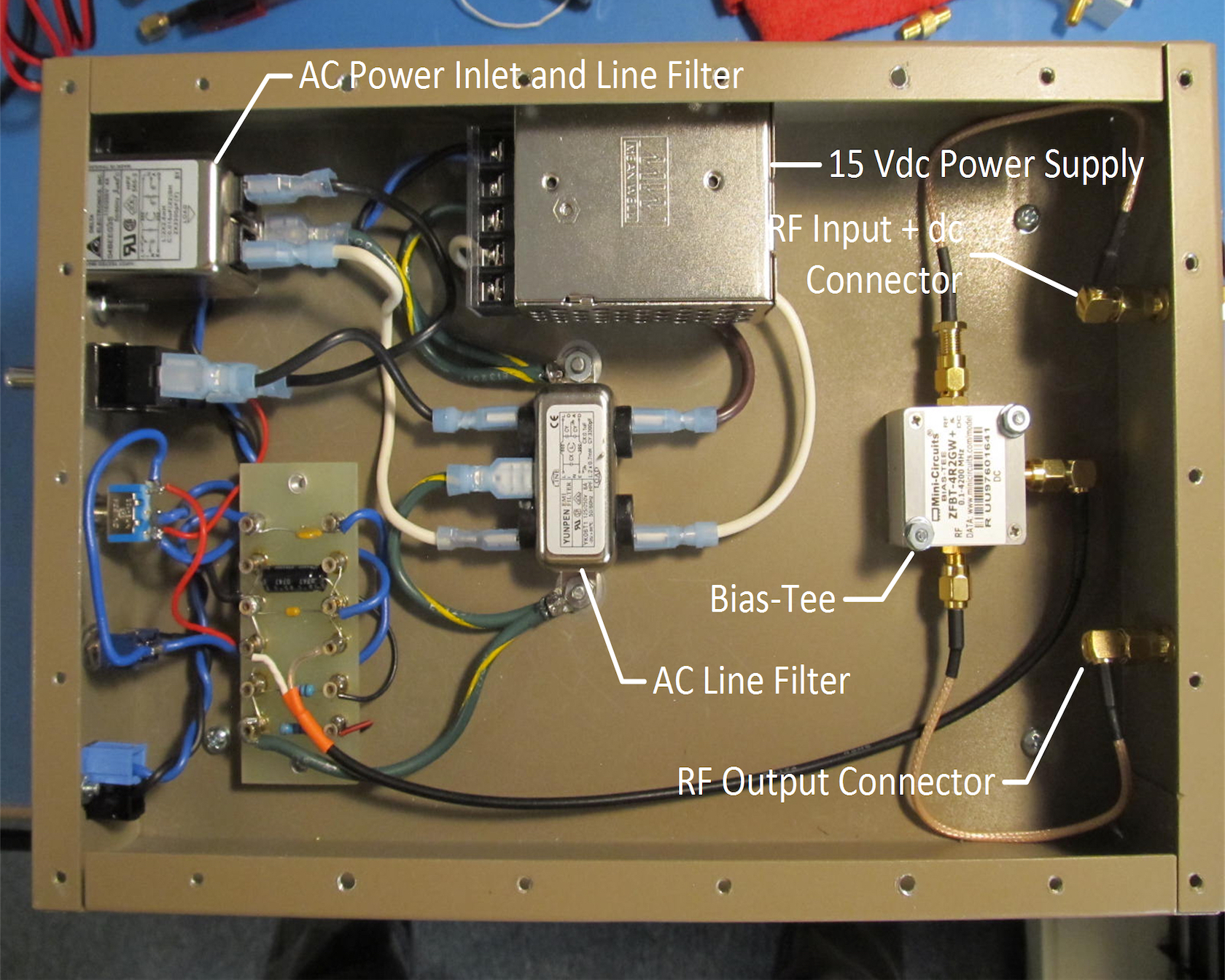}
    }%
    \caption{FEE Test Fixture box which was fabricated to 
    measure the scattering parameters of the LWA FEE board. It consists of 
    two bays with the upper bay containing a Coupler PCB which 
    converts the unbalanced signal from the VNA to a balanced 
    signal which can be injected into the FEE feedpoints. The lower
    bay contains a 15 Vdc power supply which powers the FEE--under--test
    through a bias tee.}
    \label{fig: FEE_test_fixture}
\end{figure}

The scattering parameter measurements of the FEE were carried out in a fashion similar to that of the antennas. A custom FEE 
Test Fixture box was fabricated that runs the signal from the VNA through a TeleTech HX62A hybrid coupler in reverse so the signal 
can be converted from the unbalanced coaxial cables on the VNA to a balanced signal which is injected into the FEE--under--test. 
It consists of two bays, an upper bay and a lower bay, with the FEE mounted in the upper bay on a Coupler PCB which has a hybrid 
coupler. The lower bay consists of a 15 Vdc power supply which powers the FEE through a bias tee. Labelled photos of the FEE Test 
Fixture and the upper and lower bays can be seen in Figure \ref{fig: FEE_test_fixture}. This custom mount allows for measurements 
of a single polarization of the FEE board, but the cables can be easily switched to measure the second polarization. Two port 
measurements of the FEE board also allow for updated measurements of the FEE forward gain through measurement of the FEE's S21 
parameter. This is also used as a correction to the data in the LWA1 Low Frequency Sky Survey and is therefore also of great interest.

\begin{figure}
    \centering
    \includegraphics[width=\textwidth]{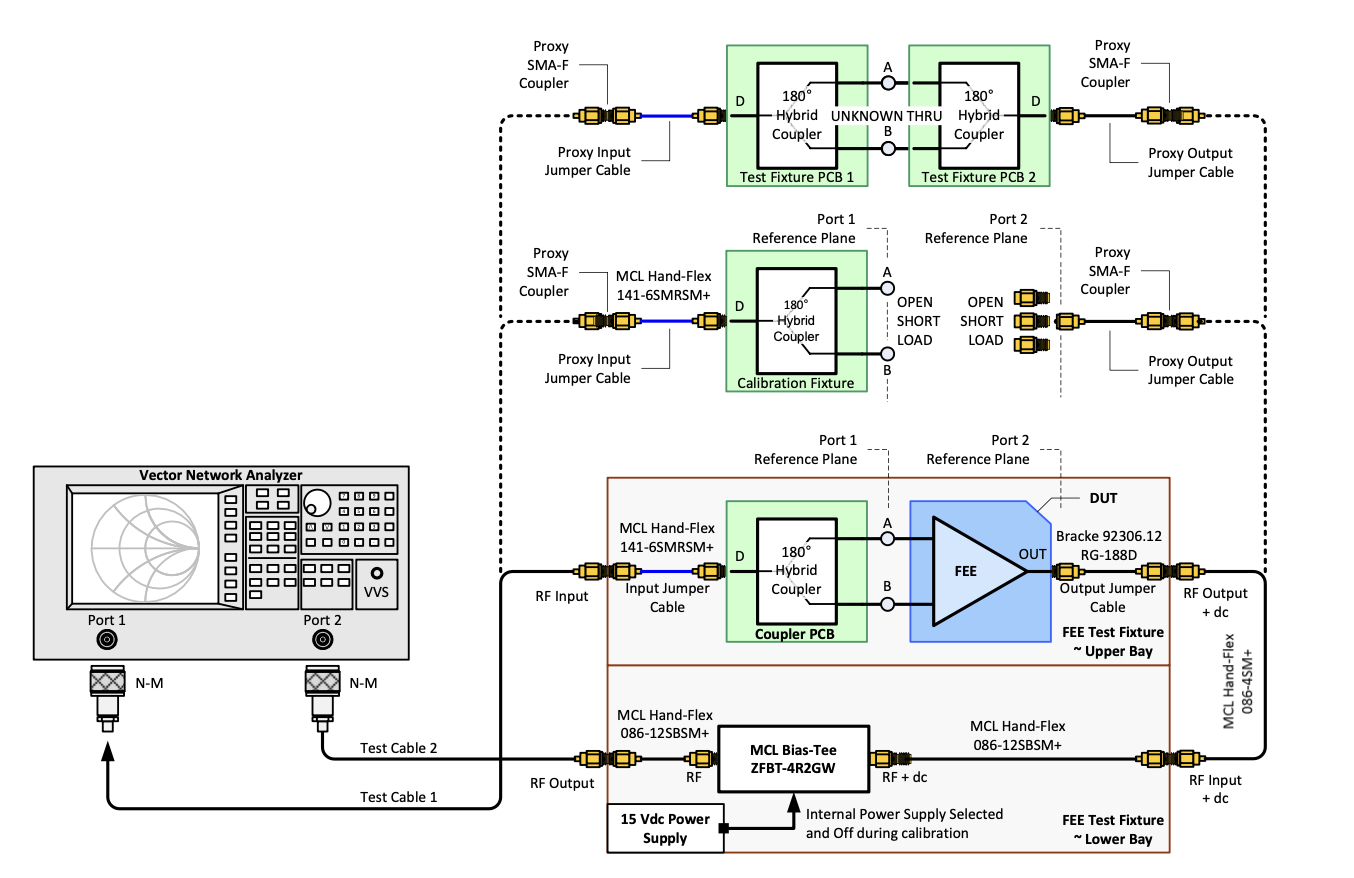}
    \caption{LWA FEE Test Fixture box. The FEE Test Fixture box consists of two bays where the FEE--under--test is mounted in the upper bay 
    and a bias tee is located in the lower bay. A $180\degree$ hybrid coupler is connected to the VNA in the upper bay in reverse in order 
    to convert the unbalanced signal from the VNA to a balanced signal which can be injected into the FEE. This allows for 2--port scattering 
    parameter measurements of the LWA FEE. The hyrbid coupler was de--embedded by using the Calibration Fixtures described in Section 
    \ref{sec: fixtures}. This procedure is represented in the plot by the setups depicted with dotted lines. Proxy cables were used to 
    account for the connections internal to the FEE Test Fixture box. The top path shows the Thru calibration, the middle path shows the 
    Short, Open, and Load calibrations, and the bottom path with solid lines show the measurement setup.}
    \label{fig: fee_schematic}
\end{figure}

However, the presence of the hybrid coupler connecting the VNA to the FEE--under--test means that the reference plane of the VNA is no longer 
at the feedpoints of the FEE board, but rather it is where the coaxial cable connects to the hybrid coupler. Calibration of the VNA using the 
custom Calibration Fixtures described in Section \ref{sec: fixtures} was carried out in order to de--embed the hybrid coupler. The calibration 
was achieved by connecting the VNA to proxy cables which account for the connections which are internal to the FEE Test Fixture box and then 
connecting the Short, Open, Load, and Thru Calibration Fixtures to Port 1 on the VNA. Port 2 was calibrated using standard SMA--F 
calibration standards. This 2--port calibration once again shifts the reference plane of Port 1 to the feedpoints of the FEE where the 
board mounts to the antenna and that of Port 2 to output of the FEE. A schematic view of the calibration and measurement procedures is 
shown in Figure \ref{fig: fee_schematic}.

\begin{figure}
    \centering
    \subfloat[Polarization A]{%
    \includegraphics[height=0.48\textheight]{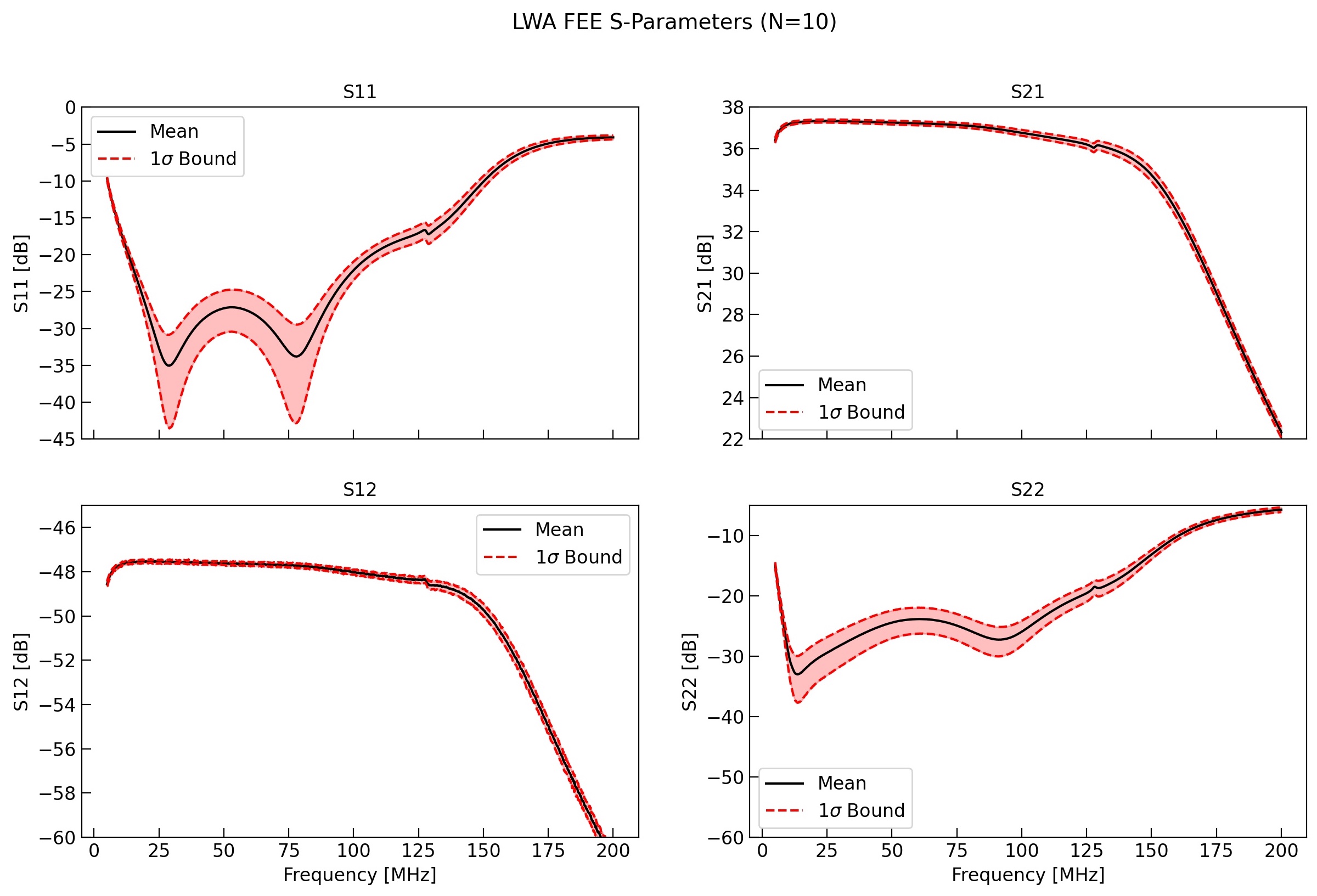}%
    } \\
    \subfloat[Polarization B]{%
    \includegraphics[height=0.48\textheight]{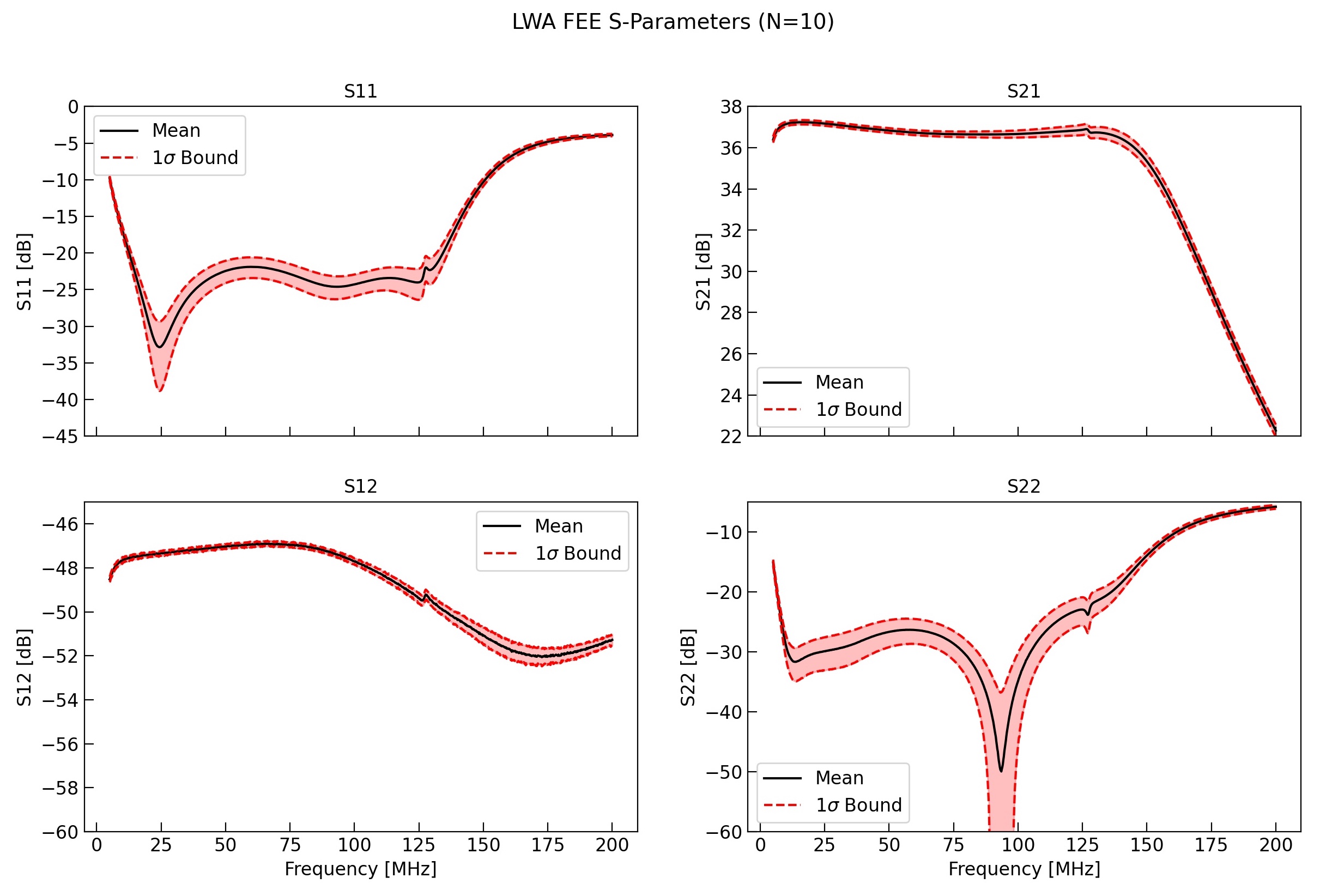}
    }
    \caption{Scattering parameters of the version 1.8 LWA FEE board. 
    Ten FEE boards were tested and the mean (black line) with $1\sigma$ 
    bounds (red shaded area) are shown.}
    \label{fig:fee_s_params}
\end{figure}

\begin{figure}
    \centering
    \subfloat[Polarization A]{%
    \includegraphics[height=0.48\textheight]{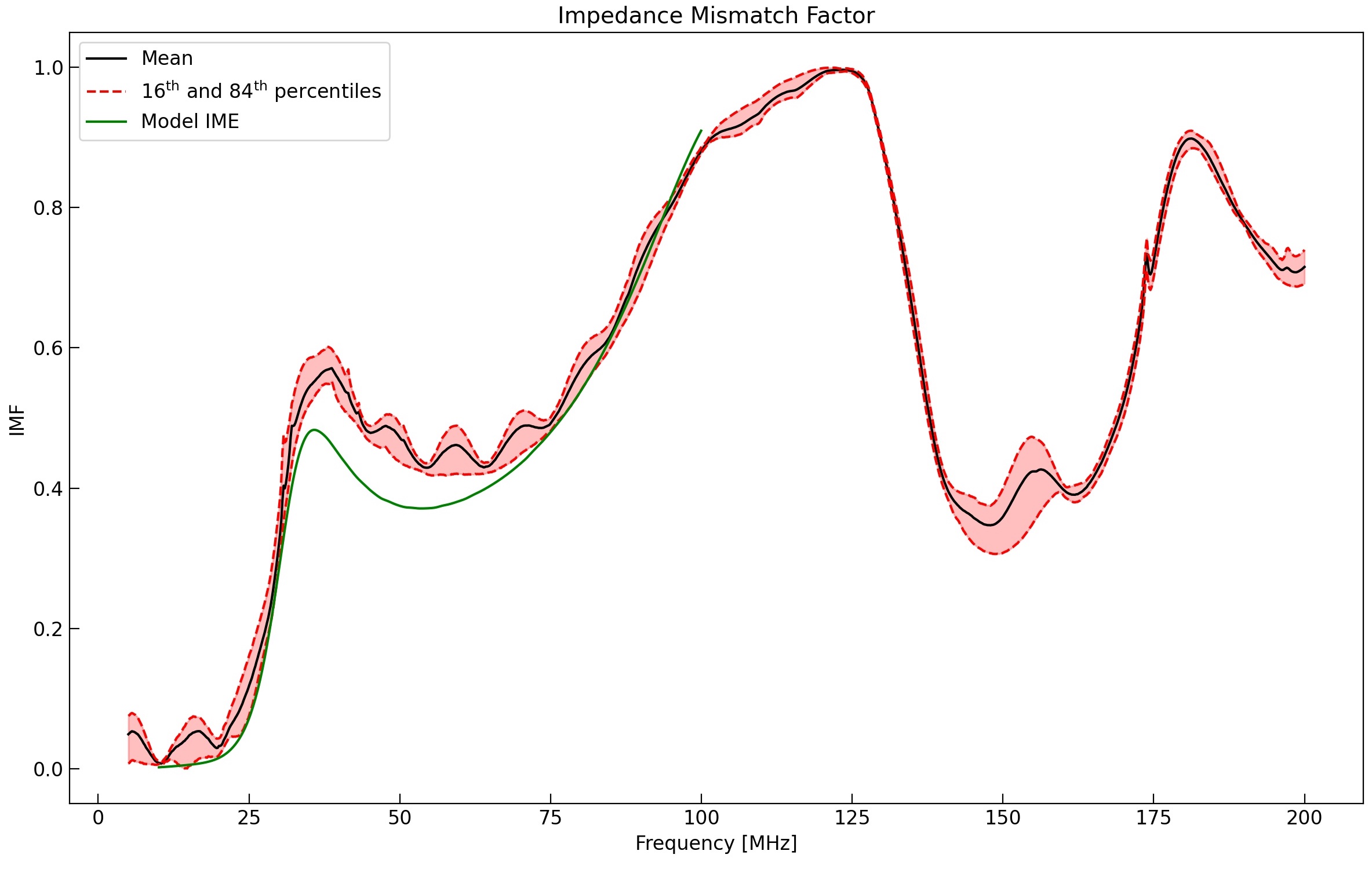}%
    } \\
    \subfloat[Polarization B]{%
    \includegraphics[height=0.48\textheight]{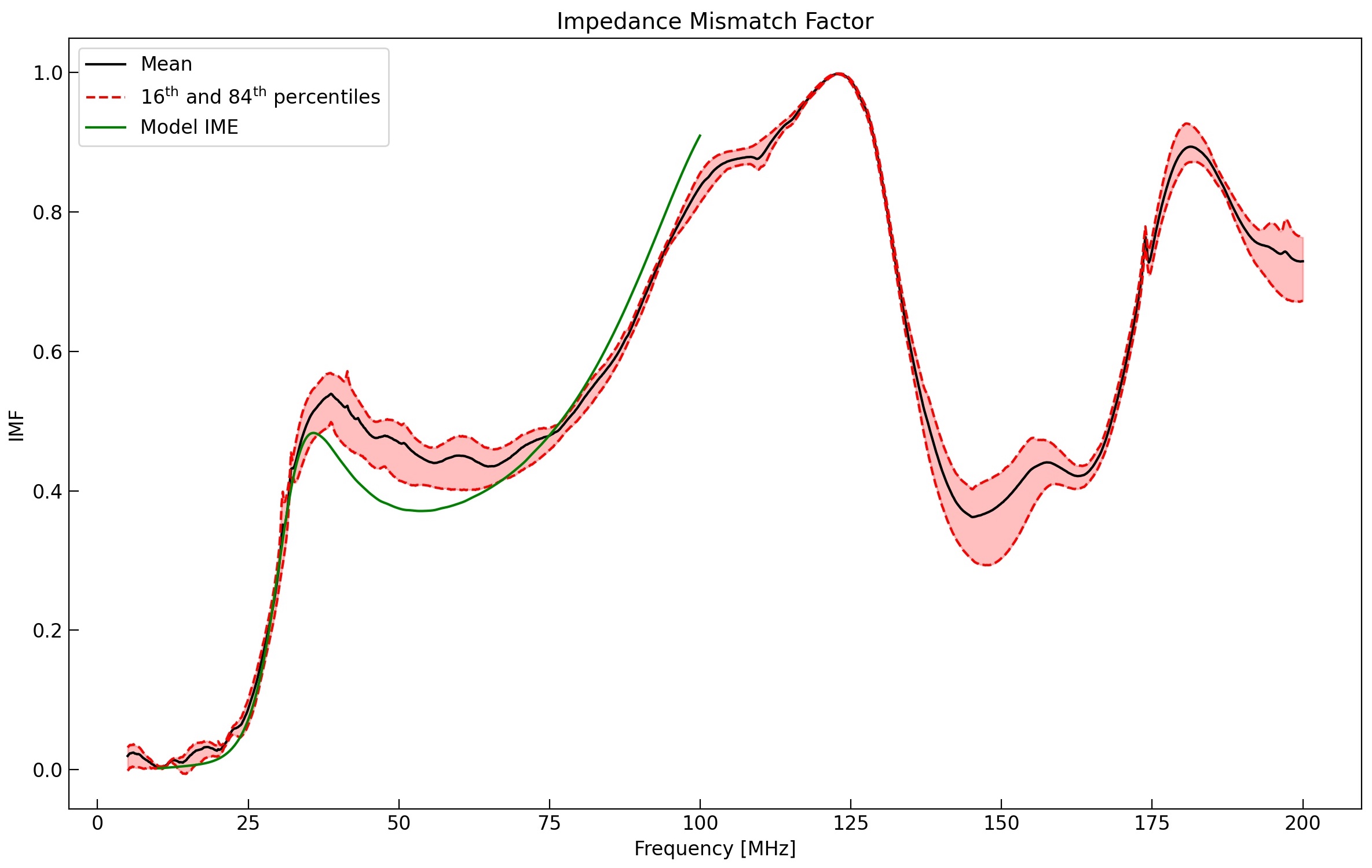}%
    }
    \caption{Impedance mismatch factor. The black curve is an average over IMFs calculated using 
    different FEE and antenna data. The red shaded area shows the $\sim 1\sigma$ uncertainty bound.
    The model IME derived in \citet{hicks2012} is shown in green.}
    \label{fig:IMF}
\end{figure}

A set of 10 version 1.8 FEE boards was used in order to get better statistics compared to a single measurement. Slight variances in components,
manufacturing, or repairs to the boards can result in changes to the scattering parameters, so it is better to compute the average 
scattering parameters from a set of FEEs. While 10 is not a very large sample, it allows for simple statistics like mean and standard 
deviation of the scattering parameters to be measured. The complete set of scattering parameters is shown in Figure \ref{fig:fee_s_params}. 
The S11 measurements were then used in combination with the single antenna measurements to compute the IMF via Equation \ref{eq: imf}. 
Antenna 162 at LWA--SV was omitted since it shows features in S11 which do not agree with the other measured antennas. 
This yields a total set of 30 IMF curves. The average of these is computed and the $16^{th}$-- and $84^{th}$-- percentiles for each 
frequency are reported to give $\sim 1\sigma$ uncertainty bounds. The measured scattering parameters of the FEE board and the computed 
IMF are shown in Figures \ref{fig:fee_s_params} and \ref{fig:IMF}, respectively.

\section{Discussion and Conclusions} \label{sec: discussion}
It is apparent from Figure \ref{fig:IMF} that the model IME presented in \citet{hicks2012} generally 
agrees with the measurements presented in this work. The new measurements imply that the antenna 
is generally more efficient than previous simulations suggested. In order to get a sense about 
how much better the new IMF captures impedance mismatch effects in the array, we use it 
to correct LWA1 Low Frequency Sky Survey and compare to results using the older IME correction. 
We also included the updated FEE forward gain correction, which is equivalent to S21 shown in Figure \ref{fig:fee_s_params}.

We used data captured using LWA1 on January $20^{\rm{th}}$, 2019 and corrected the data for impedance 
mismatch effects, FEE gain, and analog receiver board (ARX) gain. These are the three primary corrections used in \citet{dowell2017}.
In order to gauge whether the corrections improve the quality of the measured sky spectrum, we simulate the 
LWA dipole beam pattern and convolve it with a realization of the Global Sky Model \citep[GSM,][]{deOliveira2008}
at the relevant local sidereal times in order to create a simulated spectrum of the sky. The scales of the raw and 
simulated spectra must be matched since the observed raw spectrum is in an arbitrary power scale and the simulated 
spectrum is in units of temperature. The ratio of the median of simulated temperature spectrum to that of raw data spectrum
is used as a scaling factor to convert the raw data to temperature. A simple power law of the form
\begin{equation}
    T = k \cdot \nu^{\alpha} ,
\end{equation}
is then fitted to the spectrum and the spectral 
index, $\alpha$, is reported as a first order comparison to the GSM spectrum.
It should be noted that this method of assessing the quality of the corrected spectrum is not without caveats. First, there is 
debate over how accurate the GSM is at low frequencies since it relies on a principle component analysis which uses 
various input sky maps, most of which do not include frequencies below 100 MHz. Second, there is some uncertainty in 
how accurate the models of the LWA dipole beam pattern are. Together, the uncertainties in both the GSM and the LWA 
dipole beam pattern will be present in the simulated spectrum which we take to be the benchmark with which to measure 
how good our corrections are. This is not desirable, but there are no better alternatives at the present time.

\begin{figure}
    \centering
    \includegraphics[width=\textwidth]{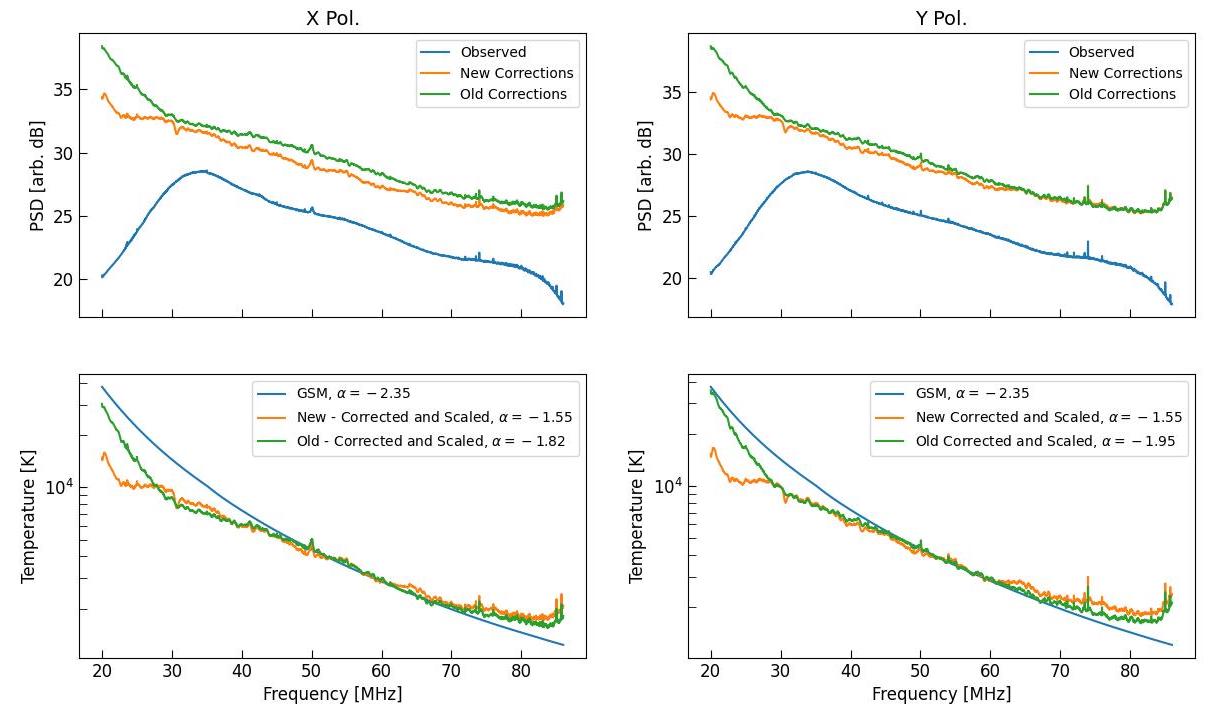}
    \caption{Comparison between old and new corrections to the LWA1 Low Frequency Sky Survey.
    The top panels show the raw observed spectra and corrected spectra. The new corrections include 
    the new IMF measurements and new FEE forward gain measurements presented in this work combined with older 
    measurements of the analog receiver (ARX) gain. The older corrections use the original IME simulations from \citet{hicks2012} 
    and previous lab--based measurements of the FEE forward gain. The bottom panels show a realization of the 
    GSM convolved with a model of the LWA dipole beam pattern along with the corrected spectra from the top panels after 
    they have been scaled to the median temperature of the GSM spectrum. A simple power law is fit for each spectrum
    and the best fit spectral index for each is reported in the legend. X polarization corresponds to the 
    dipole arms which are oriented North--South and is denoted Polarization A in other sections of this work.
    Y polarization similarly corresponds to the East--West dipole arms and is denoted Polarization B in other 
    sections of this work.}
    \label{fig:LFSS_comparison}
\end{figure}

The results of using both the old and new corrections are shown in Figure \ref{fig:LFSS_comparison}. 
The biggest difference is the removal of the severe bend in the spectrum which is present in the 
corrected spectrum that uses the older corrections at frequencies $\nu \lesssim 45$ MHz. This feature 
is not expected to be physical in the sky spectrum since the sky at these frequencies is known to be well 
modeled by a simple power law. Therefore, it is thought to be due to errors in the instrumental correction terms. 
It is encouraging to see the new correction terms reduce this feature in the corrected spectra.

The analysis presented here shows that the new IMF and FEE forward gain corrections result in a measured 
sky spectrum which is more consistent with physical expectations. However, the spectral index of the power 
law fit is still off from the expected value of the sky at these frequencies, which is $\alpha \approx -2.5$.
After updating the impedance mismatch and FEE forward gain corrections, we conclude that the final correction 
term, the ARX gain term, should be remeasured in follow up work. The current ARX correction is from old measurements
carried out in the lab which we are confident can be redone with higher accuracy and a more sophisticated methodology.
Obtaining new measurements of the ARX boards is left as future work. The above three corrections
do not account for the contribution from the receiver temperature, which is additive in nature and is expected to be 
frequency dependent. This must also be accounted for in order to fully calibrate the measured spectrum; however, this 
is beyond the scope of the presented work.

\section{Summary} \label{sec: summary}
We have carried out 2--port scattering parameter measurements of the LWA antenna and front end electronics (FEE) which 
can be used to derive correction terms to calibrate LWA data. This had not been done in the past since 
the reference plane of the VNA must be shifted to the feedpoints where the FEE boards mount to the antenna 
and the presence of a $180 \degree$ hybrid coupler on the FEE board prevented this. We developed custom 
Calibration and Test Fixtures which de--embed the hybrid coupler and shift the VNA reference plane to the antenna 
feedpoints.

We measured multiple antennas at both commissioned LWA stations in New Mexico, LWA1 and LWA--SV, as well as an isolated 
antenna located at the site of the future LWA mini--station which will be located at the end of the North arm of the Very 
Large Array, LWA--NA. We carried out both single antenna measurements which measure the four associated scattering parameters
as well as antenna--antenna coupling measurements which measure the strength of mutual coupling between adjacent elements in 
the array. Mutual coupling is very difficult to quantify in an array with a large number of elements like the LWA, which 
makes these the first measurements to ever try to quantify the amount of mutual coupling in a LWA station.

We also measured the scattering parameters for a small sample of FEE boards in order to get updated measurements on the FEE
reflection coefficient and forward gain. The FEE reflection coefficient measurements, combined with the single antenna reflection
coefficient measurements, yielded an updated impedance mismatch factor (IMF) correction used to remove impedance mismatch effects
from LWA data. We verified both the new IMF and FEE forward gain corrections by comparing their effects on LWA1 data captured in 
January of 2019. We found that the new corrections resulted in the removal of a non--physical feature in the measured sky spectrum 
which implies that they better represent the physical system than previous simulations. We conclude that we also need to make 
improved measurements of the analog receiver board (ARX) gain to fully capture instrumental effects for future sky surveys with 
the LWA.

\begin{acknowledgements}

Christopher DiLullo’s research was supported by an appointment to the NASA Postdoctoral Program at 
NASA Goddard Space Flight Center, administered by Oak Ridge Associated Universities under contract 
with NASA.

Construction of the LWA
has been supported by the Office of Naval Research under Contract N00014-07-C-0147 and by the
AFOSR. Support for operations and continuing development of the LWA1 is provided by the Air Force
Research Laboratory and the National Science Foundation under grants AST-1835400 and AGS-
1708855.
\end{acknowledgements}

\bibliography{References}

\begin{thebibliography}{}
\expandafter\ifx\csname natexlab\endcsname\relax\def\natexlab#1{#1}\fi
\providecommand{\url}[1]{\href{#1}{#1}}
\providecommand{\dodoi}[1]{doi:~\href{http://doi.org/#1}{\nolinkurl{#1}}}
\providecommand{\doeprint}[1]{\href{http://ascl.net/#1}{\nolinkurl{http://ascl.net/#1}}}
\providecommand{\doarXiv}[1]{\href{https://arxiv.org/abs/#1}{\nolinkurl{https://arxiv.org/abs/#1}}}

\bibitem[{Cranmer {et~al.}(2017)Cranmer, Barsdell, Price, Dowell, Garsden,
  Dike, Eftekhari, Hegedus, Malins, Obenberger, {et~al.}}]{cranmer2017}
Cranmer, M.~D., Barsdell, B.~R., Price, D.~C., {et~al.} 2017, Journal of
  Astronomical Instrumentation, 6, 1750007

\bibitem[{de~Oliveira-Costa {et~al.}(2008)de~Oliveira-Costa, Tegmark, Gaensler,
  Jonas, Landecker, \& Reich}]{deOliveira2008}
de~Oliveira-Costa, A., Tegmark, M., Gaensler, B., {et~al.} 2008, Monthly
  Notices of the Royal Astronomical Society, 388, 247

\bibitem[{DiLullo {et~al.}(2021)DiLullo, Dowell, \& Taylor}]{dilullo2021}
DiLullo, C., Dowell, J., \& Taylor, G.~B. 2021, Journal of Astronomical
  Instrumentation, 10, 2150015

\bibitem[{DiLullo {et~al.}(2020)DiLullo, Taylor, \& Dowell}]{dilullo2020}
DiLullo, C., Taylor, G.~B., \& Dowell, J. 2020, Journal of Astronomical
  Instrumentation, 9, 2050008

\bibitem[{Dowell {et~al.}(2017)Dowell, Taylor, Schinzel, Kassim, \&
  Stovall}]{dowell2017}
Dowell, J., Taylor, G.~B., Schinzel, F.~K., Kassim, N.~E., \& Stovall, K. 2017,
  Monthly Notices of the Royal Astronomical Society, 469, 4537

\bibitem[{Ellingson(2011)}]{ellingson2011}
Ellingson, S.~W. 2011, IEEE Transactions on Antennas and Propagation, 59, 1855

\bibitem[{Hicks {et~al.}(2012)Hicks, Paravastu-Dalal, Stewart, Erickson, Ray,
  Kassim, Burns, Clarke, Schmitt, Craig, {et~al.}}]{hicks2012}
Hicks, B.~C., Paravastu-Dalal, N., Stewart, K.~P., {et~al.} 2012, Publications
  of the Astronomical Society of the Pacific, 124, 1090

\bibitem[{Rudge {et~al.}(1986)Rudge, Milne, \& Olver}]{alan1986}
Rudge, A.~W., Milne, K., \& Olver, A.~D. 1986, The Handbook of Antenna Design
  (P. Peregrinus)

\bibitem[{Taylor {et~al.}(2012)Taylor, Ellingson, Kassim, Craig, Dowell, Wolfe,
  Hartman, Bernardi, Clarke, Cohen, {et~al.}}]{taylor2012}
Taylor, G., Ellingson, S., Kassim, N., {et~al.} 2012, Journal of Astronomical
  Instrumentation, 1, 1250004

\end{thebibliography}
\bibliographystyle{aasjournal}

\end{document}